\documentclass[useAMS,usenatbib]{mn2e}
\usepackage{graphicx,amsmath,txfonts}

\title{Fast migration of low-mass planets in radiative discs}
\author[A. Pierens]{A. Pierens $^{1,2}$ \\
$^1$Universit\'e de Bordeaux, Observatoire Aquitain des Sciences de l'Univers,
    BP89 33271 Floirac Cedex, France \\
$^2$CNRS, Laboratoire d'Astrophysique de Bordeaux,
     BP89 33271 Floirac Cedex, France\\}
\date{Released 2012 Xxxxx XX}

\pagerange{\pageref{firstpage}--\pageref{lastpage}} \pubyear{2014}

\def\LaTeX{L\kern-.36em\raise.3ex\hbox{a}\kern-.15em
    T\kern-.1667em\lower.7ex\hbox{E}\kern-.125emX}
\begin{document}
\label{firstpage}
\maketitle
\begin{abstract} 
Low-mass planets are known to undergo Type I migration and this process 
must have played a key role during the evolution of planetary systems. Analytical formulae for the disc torque have been derived assuming that the planet evolves on 
a fixed circular orbit. However, recent work has shown that in isothermal discs,  a migrating protoplanet may also experience dynamical corotation torques that scale 
with the planet drift rate.  The aim of this study 
is to examine whether dynamical corotation torques can also affect the migration of low-mass planets in non-isothermal discs. We performed 
2D radiative hydrodynamical simulations to  examine the orbital evolution outcome of migrating protoplanets as a function of  disc mass. 
 We find that a protoplanet can enter a fast migration regime when it migrates in the direction set by the entropy-related 
horseshoe drag and when the Toomre stability parameter is less than a threshold value below which the horseshoe region contracts into a tadpole-like 
region. In that case, an underdense trapped region appears near the planet, with an entropy excess compared to the ambient disc.  If the viscosity and thermal diffusivity are small enough so that the entropy excess is conserved during migration, the planet then experiences strong corotation 
torques arising from the material flowing across the planet orbit.
During fast migration, we observe that a protoplanet can pass through the zero-torque line predicted by static torques.  We also find that fast migration may help in disrupting the mean-motion resonances that are formed by convergent migration of embryos.
\end{abstract}

\section{Introduction}
A striking feature of the population of exoplanets discovered so far is their great diversity. Compact and non-resonant 
systems of super-Earths and  mini-Neptunes have been discovered orbiting within a few tenths of AU from their stars 
(e.g. Lissauer et al. 2011; Lovis et al. 2011). Giant 
planets with periods from days to  several thousands of days have also  been found around $\sim 14 \%$ of Sun-like stars, 
with a frequency that is strongly correlated to the metallicity of the host star (Cumming et al. 2008; Mayor et al. 2011).  A main aim of planet 
formation and evolution  scenarii is to examine whether this broad diversity  can be explained, and under which conditions.  
These scenarii can be constrained using for example planet population syntheses which combine analytical models of disc evolution,  
planet formation and orbital evolution due to  disc-planet interactions (Ida \& Lin 2008; Mordasini et al. 2009; Mordasini et al. 2012); and generate a large  variety of system architectures that 
can be statistically compared to observations. An alternative method  is to make use of N-body simulations of planetary 
systems formation which do 
not cover a parameter space as large as in planet population syntheses, but allow for  a self-consistent treatment
of planet-planet interactions (Hellary \& Nelson 2012; Coleman \& Nelson 2014; Cossou et al. 2014). 

In both of these approaches, disc-planet interactions causing planet 
migration are modeled using analytical formulae for the disc torque experienced by the planet. 
This gravitational  torque consists of two components. The differential
Lindblad torque results from the angular momentum exchange between the planet and the spiral density waves it
  generates inside the disc.  For sufficiently low-mass planets, it  scales linearly with disc mass and planet mass and as the inverse square of the disc aspect ratio (Tanaka et al. 2002). Although its sign depends on the  density and temperature gradients inside the  disc, the differential Lindblad
torque is generally negative for typical disc models and is therefore responsible for inward migration.
The corotation torque is due to the torque exerted by the material located in the coorbital region of the planet. It is 
composed  of a barotropic part  which scales with the vortensity (i.e.
the ratio between the vertical component of the vorticity and the disc surface density) gradient
 (Goldreich \& Tremaine 1979) plus an entropy-related part which scales with the entropy gradient (Baruteau \& Masset 2008; Paardekooper \& Papaloizou 2008).
A negative vortensity (resp. entropy) gradient gives
rise to a positive vortensity (resp. entropy) related corotation torque. It has been shown that for midly
positive surface density gradients or negative entropy gradients, a positive  corotation torque can eventually counteract the effect of a
negative differential Lindblad torque,  which may stall or even reverse migration (Masset et al. 2006; Paardekooper \& Papaloizou 2009).
 In isothermal discs, the corotation torque is a non-linear process generally referred as the horsehoe drag and whose amplitude
is controlled by advection and diffusion of vortensity inside the horsehoe region. In non-isothermal discs, the 
corotation torque is also powered by singular production
of vortensity due to an entropy discontinuity on downstream separatrices (Masset \& Casoli 2009;
Paardekooper et al. 2010).
In the absence of any diffusion processes inside the disc, vortensity and entropy gradients
 across the horseshoe region tend to flatten through phase mixing, which causes the two components of the
horseshoe drag to  saturate.
Consequently, desaturating the horseshoe drag requires that some amount of viscous and thermal diffusions are operating inside the
horseshoe region. In that case, the amplitude of the horseshoe drag depends on the ratio between the diffusion timescales and the horseshoe
libration timescale and its optimal value, also referred as the fully unsaturated horseshoe drag, 
is obtained when the diffusion timescales are approximately equal to half the horseshoe libration time
 (e.g. Baruteau \& Masset 2013).
In the limit where the diffusion timescales become shorter than the U-turn timescale, the corotation torque decreases and approaches the value
predicted by linear theory. Therefore, the corotation torque can be considered as a linear combination of the fully unsaturated
horseshoe drag and the linear corotation torque with coefficients depending on the ratio between the diffusion timescales and
the horseshoe libration timescale. Corotation torque
formulae as a function of viscosity and thermal diffusivity were recently proposed by Paardekooper et al. (2011) and
Masset \& Casoli (2010). 

Such analytical formulae have been derived assuming that the planet evolves on a fixed circular orbit. In that case, the 
planet feels a static torque that depends only on the local temperature and surface density conditions in the disc.  If the planet
is allowed to migrate, however, 
the disc torque can also exhibit, under certain conditions,  a dependence on the planet drift rate.  
For instance, this can occur if the planet is massive enough to partly deplete  its coorbital region. In that case,  a coorbital mass deficit is 
created and the corotation torque is essentially due to the gas material that flows across the planet's 
horseshoe region as the latter migrates. This contribution to the corotation torque  not only scales with the planet drift rate, but also has the same sign as the 
drift rate. This yields a positive feedback on migration, and eventually gives rise to runaway effects if the disc is massive enough (Masset \& Papaloizou 2003). 

In a recent study, Paardekooper (2014) has  shown that low-mass planets embedded in isothermal discs a few times more massive than the 
Minimum Mass Solar Nebula (hereafter MMSN; Hayashi 1981) may  also experience runaway migration in some cases. For low-mass planets that do not significantly perturb the underlying disc structure, this process does not result from  a coorbital mass deficit but rather arises due to the action of dynamical  
corotation torques  that scale with the planet drift rate and depend on  
the vortensity gradient inside the disc (Paardekooper 2014). In the case where the migration proceeds in the direction set by the corotation torque, 
Paardekooper (2014) has shown that these dynamical torques have a positive feedback on migration,  and can possibly give rise to high drift rates.  Interestingly, using an 
isothermal disc model with a positive surface density gradient to mimic what may happen in non-isothermal disc, Paardekooper (2014) has shown that an outward migrating 
protoplanet subject to strong dynamical torques  can pass through the location of the zero-torque radius where the Lindblad torque and the 
(static) corotation torque cancel each other. 

 In this paper, we focus on  the case of non-isothermal disc models, in which outward migration can 
 proceed due to the operation of  the entropy-related horseshoe drag. The aim of this study is to examine the plausibility that 
low-mass planets embedded in non-isothermal discs experience strong dynamical corotation torques as they migrate.  Using hydrodynamical simulations, we find 
that in relatively massive discs with an initial large-scale entropy gradient, the drift rates reached by a migrating planet can indeed be significantly higher than those expected by assuming 
a static torque.  We show that this occurs when i) the planet migrates in the direction set by the entropy-related horseshoe drag, and ii) the Toomre stability parameter 
is less than a critical value that depends on the entropy gradient.  Below this value, the horseshoe region does not extend to the full $2\pi$ in 
azimuth and an underdense trapped region appears which causes the planet to undergo large 
drift rates. In this fast migration regime, we also find that an outward migrating protoplanet can migrate well beyond the location of the zero-torque radius, 
similarly to the findings of Paardekooper (2014) in isothermal discs.

This paper is organized as follows. In Sect. 1, we present the numerical setup. In Sect. 2, we present the results of the simulations for non-isothermal 
disc models. The case of radiative disc models that include the effects of viscous plus stellar heating, and radiative cooling are discussed in Sect. 3. 
Finally,  we discuss our results and draw our conclusions in Sect. 4.

\section{The hydrodynamical model}
Simulations were performed using the GENESIS (De Val-Borro et al. 2006) numerical code which solves 
the equations governing the disc evolution on a polar grid $(R,\phi)$ using an advection scheme based on the monotonic  transport 
algorithm (Van Leer 1977). It uses the FARGO algorithm (Masset 2000) to avoid time step limitation due to the 
Keplerian velocity at the inner edge of the disc.  

We adopt computational units such that the mass of the central star is $M_*=1M_\odot$, the gravitational constant is 
$G=1$, and the radius $R=1$ in the computational domain corresponds to $5.2$ AU.  When discussing the results of the simulations, 
 time will be measured in units of  the orbital period at $R=1$.
 
We do not consider the disc self-gravity in this work. The  typical value for  the Tooomre stability parameter ${\cal Q}= \kappa c_s/\pi G \Sigma$, 
where $\kappa$ is the epicyclic frequency, $c_s$ the sound speed and $\Sigma$ the surface density is ${\cal Q}\sim 3-4$, suggesting thereby 
that the effect of self-gravity may be important. 
The gravitational potential for the disc therefore restricts to the gravitational 
potential from the star and the planet, plus the indirect terms that account for the fact that the frame centered on the 
star is not inertial. The gravitational 
potential of the planet is smoothed over a distance $b=0.4H_p$, where $H_p$ is the disc scale height at the location of the planet. 
Here, the planet  orbit can evolve due to the gravitational forces exerted by the star and the disc, and eventually due to the 
gravitational 
interaction of the planet with other bodies (see Sect. \ref{sec:4planets}). When calculating the disc force experienced by the planet, we follow Paardekooper (2014) and 
exclude the axisymmetric component of the force to ensure consistency as the planet migrates.

In the following, two different forms of the energy equation will be employed. We will first consider non-isothermal disc models for which 
the energy equation corresponds to an advection-diffusion equation for the gas entropy with a constant 
thermal diffusion coefficient (Paardekooper et al. 2011). Then we will focus 
on more realistic, radiative disc models whose thermal structure results from the balance between viscous plus stellar heating, and radiative cooling.
\subsection{Non-isothermal discs}
\subsubsection{Numerical setup}
For non-isothermal disc models,  we solve the following energy equation:
\begin{equation}
\frac{\partial e}{\partial t}+\nabla \cdot (e{\bf v})=-(\gamma-1)e{\nabla \cdot {\bf v}}+\chi e \nabla^2 \text{log}S
\end{equation}
where $e$ is the thermal energy density, $\bf v$ the velocity, $\gamma$ the adiabatic index which is set 
to $\gamma=1.4$, and where $\chi$ is a thermal diffusion coefficient. In the previous equation, $S=p/\Sigma^\gamma$ is  the gas entropy, with
$p=(\gamma-1)e$ the gas pressure.  Strictly speaking, the diffusion term in the previous equation should involve the laplacian of temperature rather 
than entropy. This form for the energy equation is used so that it corresponds to an advection-diffusion equation for the gas entropy, and is valid 
provided that the variations in temperature  occur over a  scale smaller than those corresponding to  the gas pressure (Masset \& 
Casoli 2010). Here, this condition if fulfilled since we are 
interested in low-mass planets for which the width of the horseshoe region is typically a fraction of the disc pressure scale height.
 The thermal diffusion coefficient $\chi$  is  chosen assuming a Prandtl number $P_R=\nu/\chi$ of unity, where $\nu$ is the kinematic 
 viscosity which is assumed to be constant.   In the case where diffusion of entropy is due to turbulent transport, this is consistent with 
the results  of Pierens et al. (2012) who found $P_R\sim 1.2$ in non-isothermal disc models with turbulence driven by stochastic 
forcing. In a real protoplanetary disc, however, where  radiative diffusion can possibly dominate over turbulent transport for diffusing 
entropy, we expect the Prandtl number to be of the order of unity only in the optically thick inner regions (Pierens et al. 2012), while 
 a value $P_r<1$ is expected in the 
outer, optically thin regions.

Typically $N_R=1600$ radial grid cells uniformly distributed between $R_{in}=0.4$ and 
$R_{out}=5$ are used, and $N_\phi=2130$ azimuthal grid cells.  

We employ closed boundary conditions at both the inner and outer edges of the computational domain, and make use of wave-killing zones for 
$R<0.5$ and $R>4.5$ to avoid wave reflections at the disc edges.

\subsubsection{Initial conditions for the non-isothermal runs}

The initial disc surface density is $\Sigma=\Sigma_0(R/a_0)^{-\sigma}$, where $\Sigma_0$ is the surface density at 
the initial position of the planet  $R=a_0$ 
and where the power law index 
is such that $\sigma=3/2$ by default. For such a disc surface density profile, the initial vortensity-related part of the corotation torque cancels out 
so that the corotation torque consists only of its entropy-related part. Following Paardekooper (2014), we parametrise the disc mass through the 
parameter
$$
q_d=\frac{\pi a_0^2\Sigma_0}{M_\star}, 
$$
for which we adopt  values ranging from   $q_d=0.005$  to  $q_d=0.03$.

The initial aspect ratio is $h=h_0(R/a_0)^f$ where $h_0=0.05$ and $f$ is the disc flaring index for which we consider values of $f=0.7$ and 
$f=-0.8$. This corresponds to initial temperature profiles that vary as $R^{-\beta}$  with $\beta=1-2f=-0.4$ and $\beta=2.6$ 
respectively, and to 
initial entropy profiles that vary as $R^{-\xi}$ with $\xi=\beta-(\gamma-1)\sigma=-1$ and 
$\xi=2$ respectively. 

Our reference value for the planet-to-star mass ratio is $q=10^{-5}$ (equivalent to a $3.3$ Earth mass planet in our units). 
For $h_0=0.05$, the half-width of the 
horseshoe region is $x_s\sim 1.1a_0\sqrt{q/h_0}\sim 0.016\;a_0$  (Masset et al. 2006), which means that the horseshoe 
region is typically resolved by about 12 grid cells in the radial direction.

\subsection{The radiative disc model}
\subsubsection{Numerical setup}
 In addition to these non-isothermal disc models, we have also  considered more realistic discs in which the effects of viscous and stellar heating 
 are balanced by radiative cooling. In that case, the following energy equation is adopted:
 \begin{equation}
\frac{\partial e}{\partial t}+\nabla \cdot (e{\bf v})=-(\gamma-1)e{\nabla \cdot {\bf v}}+Q^+_{vis}-Q^--2H\nabla  \cdot {\bf F}
\label{eq:energy}
\end{equation}
 where $H$ is the disc scale height. In the previous equation,  $Q^+_{vis}$ is the viscous heating term,  $Q^-=2\sigma_B T_{eff}^4$ is the local radiative cooling from the disc surfaces, where 
$\sigma_B$ is the Stephan-Boltzmann constant and $T_{eff}$ the effective temperature which is given by (Menou \& Goodman 2004):
\begin{equation}
T_{eff}^4=\frac{T^4-T_{irr}^4}{\tau_{eff}} \quad \text{with} \quad \tau_{eff}=\frac{3}{8}\tau+\frac{\sqrt{3}}{4}+\frac{1}{4\tau}
\end{equation}
Here, $T$ is the midplane temperature and $\tau=\kappa\Sigma/2$ is the vertical optical depth, where $\kappa$ is the Rosseland mean opacity which is taken from Bell \& Lin (1994). $T_{irr}$ is the irradiation temperature which is computed from the irradiation flux (Menou \& Goodman 2004):
\begin{equation}
\sigma_B T_{irr}^4=A \frac{L_\star(1-\epsilon)}{4\pi R^2}\frac{H}{R}\left(\frac{d \log H}{d\log R}-1\right)
\end{equation}
where $\epsilon=1/2$ is the disc albedo and $L_\star$  the stellar luminosity which is computed assuming a stellar radius $R_\star=1.5\;R_\odot$ and 
a stellar temperature $T_\star=4370$ K. The factor $A$ is set to  $A=1$ if $d(H/R)/dR>0$, whereas we set $A=0$ in the regions 
where  $d(H/R)/dR<0$ in order 
to take into account possible self-shadowing effects (Gunther \& Kley 2004).
In Eq. \ref{eq:energy}, ${\bf F}$ is the radiative flux which is treated in the flux-limited  diffusion approach and which 
reads (e.g. Kley \& Crida 2008):
\begin{equation}
{\bf F}=-\frac{16\sigma \lambda T^3}{\rho \kappa}\nabla T
\end{equation}
where $\rho=\Sigma/2H$ is the mid plane density and  where $\lambda$ is a flux-limiter (e.g.  Kley 1989).
\subsubsection{Initial conditions for the radiative runs}

For the radiative simulations,  we will focus on the case of  equilibrium discs with surface density  $\Sigma=\Sigma_0 (R/R_0)^{-1/2}$ and constant 
viscosity, where $\Sigma_0$ is the surface density at $R_0=1$.  Two disc models are considered, with parameters summarized in Table $1$. For Model $1$, $\Sigma_0=1.6\times 10^{-3}$ 
(equivalent to $\Sigma_0=520$ $\text{g.cm}^{-2}$  at $5.2$ AU) and $\nu=10^{-5}$, whereas for Model $2$,  $\Sigma_0=3.2\times 10^{-3}$ (equivalent to $\Sigma_0=1024$ $\text{g.cm}^{-2}$ at $5.2$ AU) and $\nu=10^{-6}$.

The resolution used in each simulation is $(N_R,N_\phi)=(1024,1536)$  with a  radial extent corresponding to 
 $R\in[R_{in}, R_{out}]$, where $R_{in}$ and $R_{out}$ are given in Table $1$. 
 
 We adopt planet masses such that $q=3\times 10^{-5}, 6\times 10^{-5}, 10^{-4}$ for Model $1$ and $q=10^{-5}, 1.5\times 10^{-5}, 2\times 10^{-5}$ 
for Model $2$.  For each model,  the planet's horseshoe region is resolved at minimum by $\sim 11$ grid cells 
along the radial direction, which leads  to a relative error on the estimation of the corotation torque of $\sim 10 \%$ (Masset 2002).

\begin{figure*}
\centering
\includegraphics[width=0.49\textwidth]{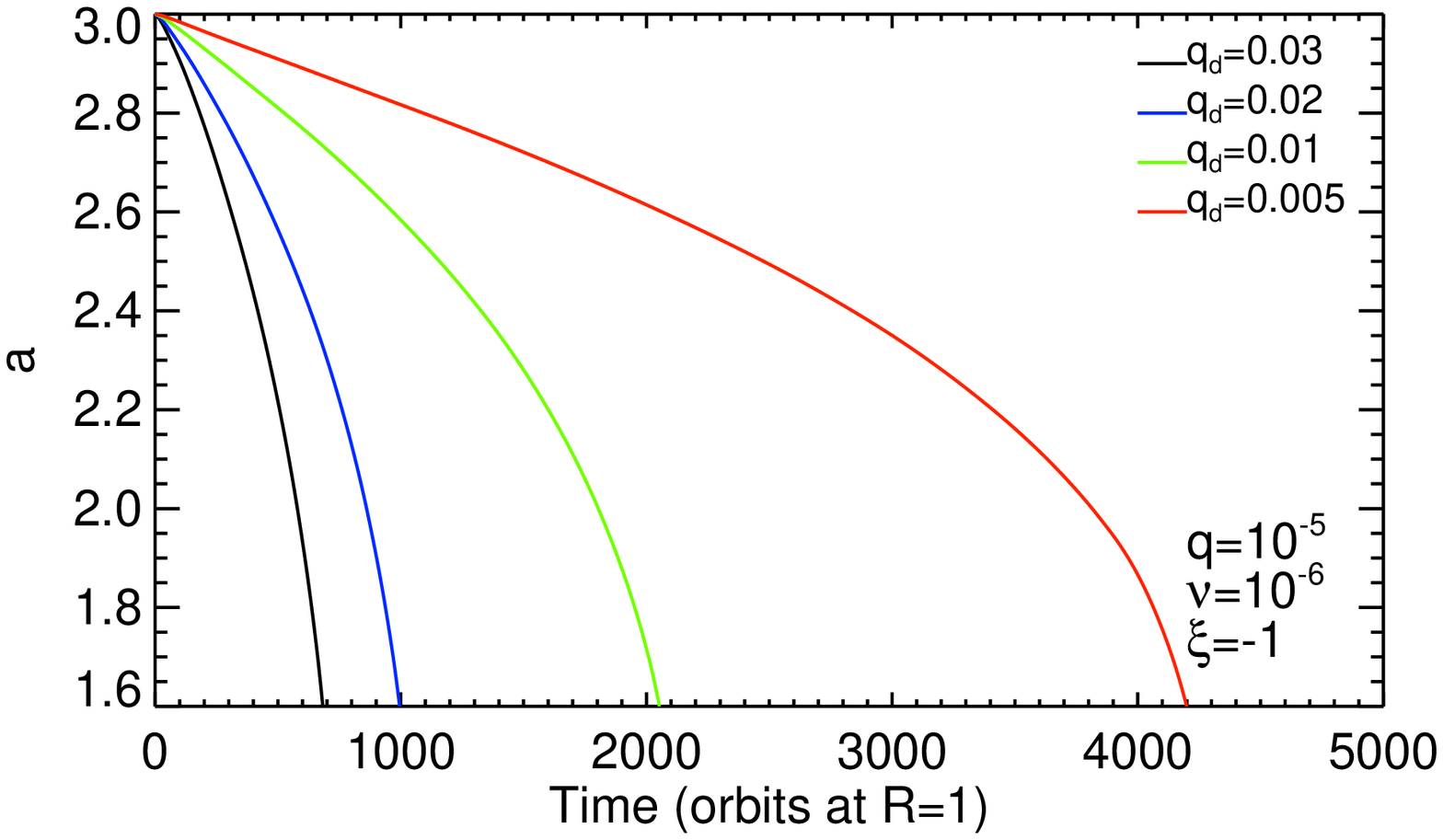}
\includegraphics[width=0.49\textwidth]{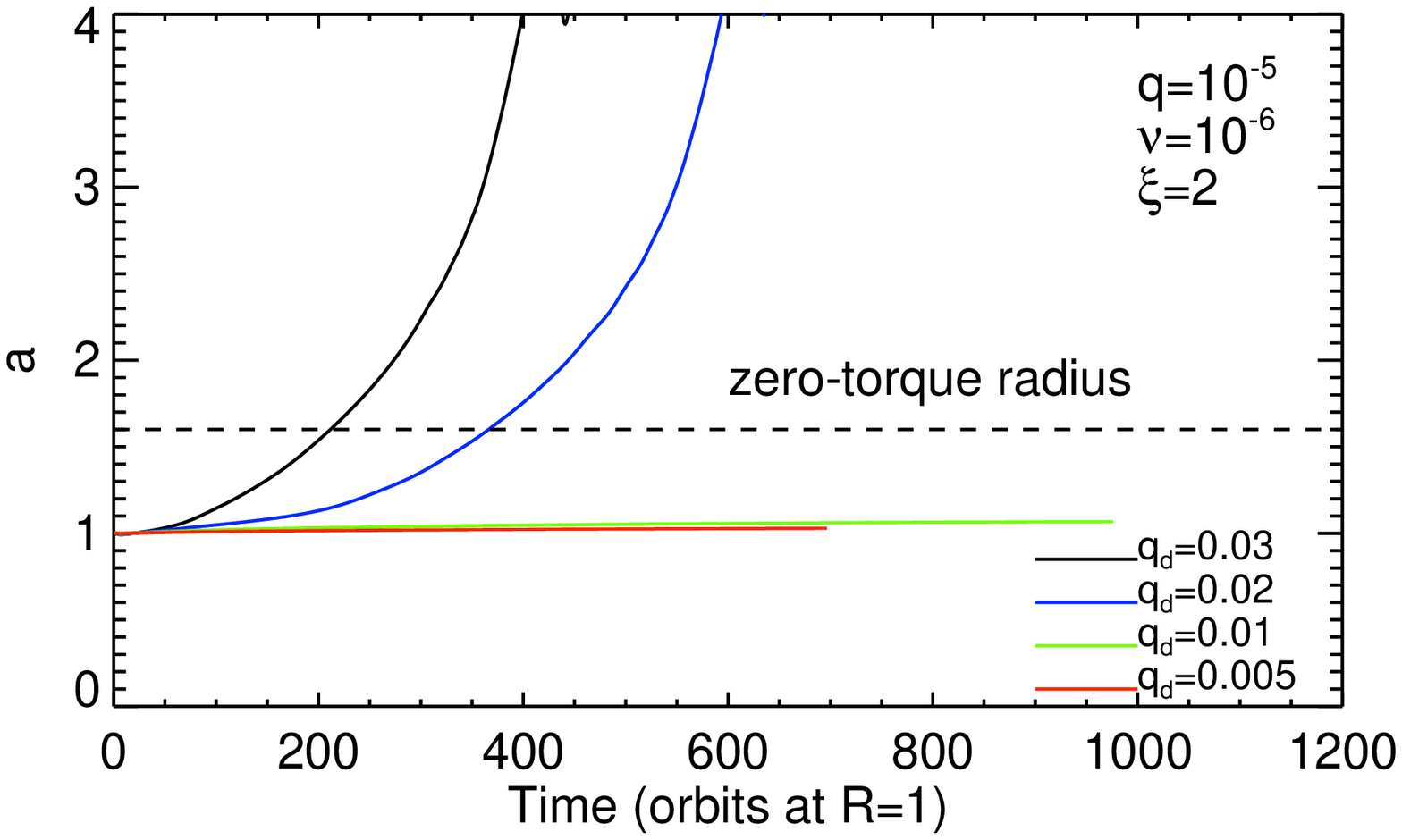}
\includegraphics[width=0.49\textwidth]{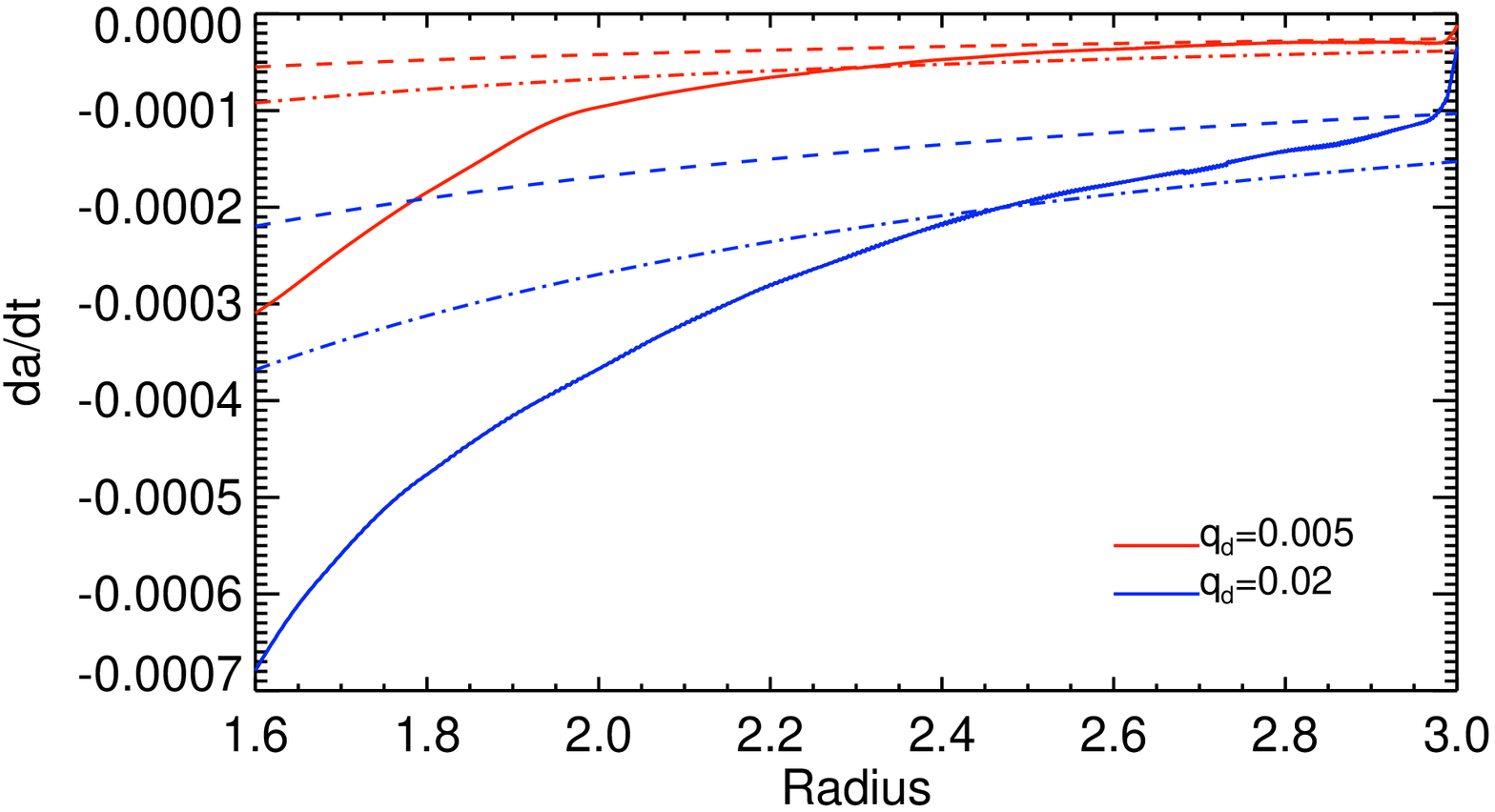}
\includegraphics[width=0.49\textwidth]{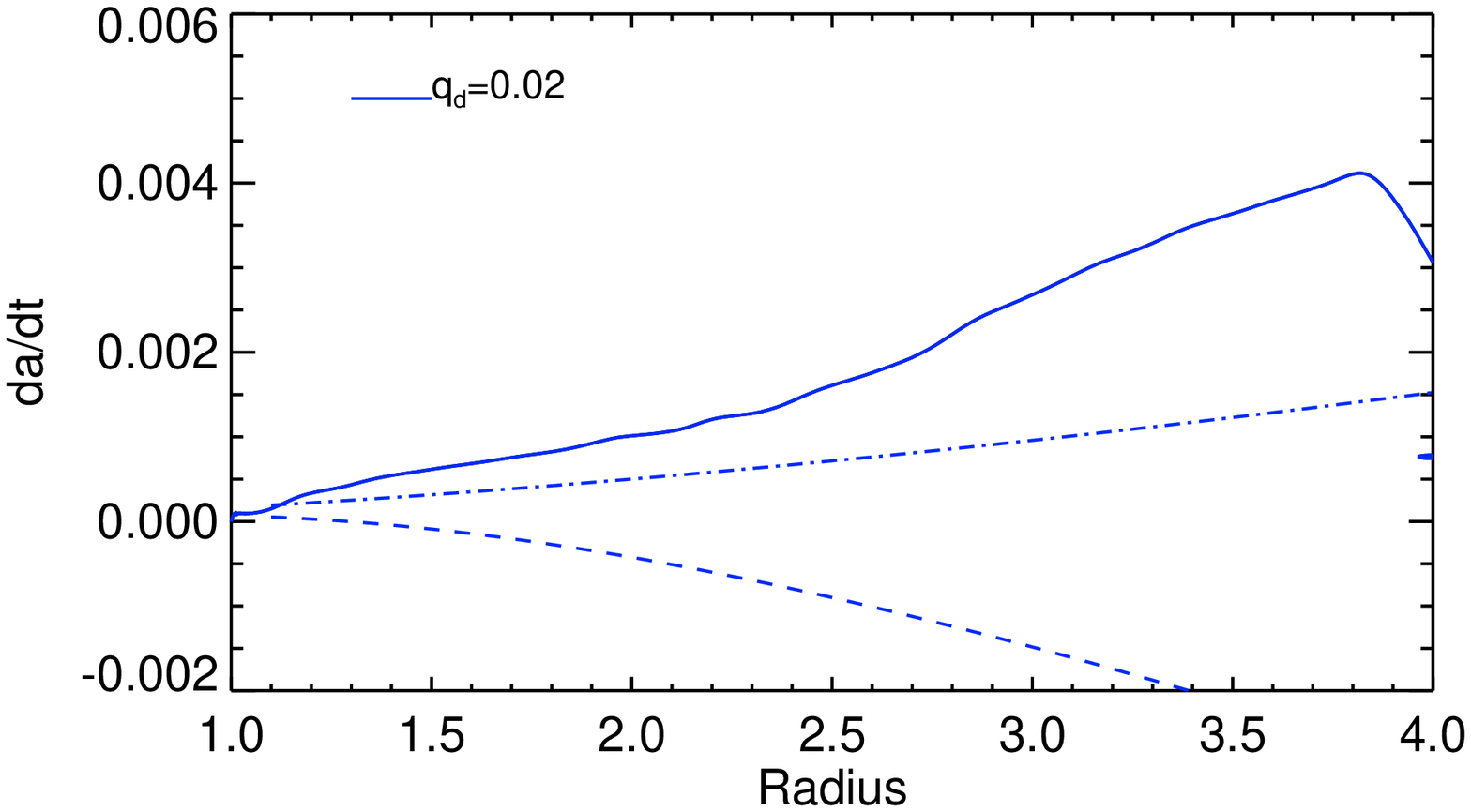}

\caption{{\it Left panel}:  Time evolution of the semi-major axis (top) and drift rate as a function of radius 
(bottom) of a protoplanet with $q=10^{-5}$ embedded in a non-isothermal disc with $\xi=-1$ and $\nu=\chi=10^{-6}$, for 
various disc masses. The dashed line shows the drift rate that is expected from analytical formulae for the disc torque, whereas the 
dot-dashed line corresponds to the drift rate expected assuming a fully unsaturated entropy-related 
horseshoe drag.  {\it Right panel:} same but for an initial entropy gradient corresponding to $\xi=2$.}
\label{fig1}
\end{figure*}

\begin{table}
\caption{Parameters of the radiative simulations, where $\Sigma=\Sigma_0(R/R_0)^{-1/2}$ and where the viscosity is constant}
\label{table}
\begin{tabular}{ccccc}
\hline
\hline
Model & $\Sigma_0$ & $\nu$ & $R_{in}$ & $R_{out}$ \\
\hline
Model $1$ & $1.6\times 10^{-3}$ & $10^{-5}$ & $0.4$ &  $6$ \\
Model $2$ & $3.2\times 10^{-3}$ & $10^{-6}$ & $0.4$ &  $3.5$ \\
\hline
\end{tabular}
\end{table}

\section{Non-isothermal disc models}

\subsection{Results }
\label{sec:results}

For a surface density profile with $\sigma=1.5$, the total disc torque $\Gamma$  exerted on a protoplanet held on a fixed circular orbit, which in the following we will refer to as the static torque, 
can be considered as the 
sum of the differential Lindblad torque,  and  the entropy-related corotation torque. In the limit where the corotation 
torque is fully unsaturated, $\Gamma$ is given by (Paardekooper et al. 2010):
\begin{equation}
\gamma \Gamma/\Gamma_0=-2.5-1.7\beta+0.1\sigma+7.9\frac{\xi}{\gamma},
\label{eq:gsurg0}
\end{equation}
with $\Gamma_0=(q/h_p)^2\Sigma_p a^4\Omega_p^2$. Here $a$ is the planet semi-major axis and the  subscript "p" indicates that 
quantities are evaluated at the orbital position 
of the planet.  For $\xi=-1$, this gives 
$\gamma \Gamma/\Gamma_0 \sim -7.3$, whereas $\gamma \Gamma/\Gamma_0 \sim 4.5$ for 
$\xi=2$. For these two disc models, the total and entropy-related corotation torques have therefore the same sign,  and the aim 
of this section is to examine whether in that case,  a migrating protoplanet can reach migration rates that are significantly higher 
than those obtained by evaluating the static torque only.  In 
isothermal discs a few times more massive than the MMSN,  it has been shown that if the  planet migrates in the direction set by the corotation torque,  it 
can be subject to strong dynamical corotation torques and runaway 
effects can even be triggered under certain conditions (Paardekooper 2014). 

The upper left panel of Fig. \ref{fig1} shows, for  the disc  model with $\xi=-1$ and for various disc masses,  the semi-major axis evolution of  a $q=10^{-5}$ planet  initially 
located at $a_0=3$. In agreement with the above estimation of the total torque, the planet migrates inward in that case,  and in  the direction set by the 
static corotation torque since the entropy gradient is positive. Here, both the viscosity and  thermal diffusivity are set to $\chi=10^{-6}$, consistently with the assumption of a Prandtl number of unity. This value  for the thermal diffusivity is chosen such that the entropy-related horseshoe drag 
 is initially close to its unsaturated value. This is reached when the thermal diffusion timescale across the horseshoe region $\tau_d=x_s^2/\chi$ 
is approximately equal to half the libration timescale $\tau_{lib}=8\pi a/(3 \Omega_p x_s)$.  However,  it can be easily shown that 
$\tau_d/\tau_{lib}\propto a^{(1-3f)/2}$,  which evaluates to  $\tau_d/\tau_{lib}\propto a^{-0.55}$ for $f=0.7$. Therefore, we expect the level of saturation of the  corotation torque to increase as the planet 
migrates inward. 
In the lower left panel of Fig. \ref{fig1} are compared, for the runs with $q_d=0.005, 0.02$,   the migration rates at different locations  with those predicted  from  static torques.  Also 
plotted on this figure are the drift rates obtained assuming a fully unsaturated horseshoe drag, and which should therefore correspond, in principle, to 
the highest drift rates that could be reached by a migrating protoplanet. 
   Clearly, the migration rates are significantly higher in the case where the planet is allowed to 
migrate, with values that appear to be even larger  than the estimate given by assuming a fully unsaturated corotation torque.


The right panel of Fig. \ref{fig1} presents the results of the simulations for the disc model with $\xi=2$ and for which outward migration is expected.  
  We note that the run with $q_d=0.03$ serves for illustrative purpose only since ${\cal Q} < 1$ for $R \gtrsim 2$ in that case.
Here $\tau_d/\tau_{lib}\propto a^{1.7}$,  leading to an increase in the saturation of the corotation torque as the planet migrates 
outward.  The main implication of this is that  the total torque acting on the planet can eventually cancel at the location where 
the differential  Lindblad torque and the saturated corotation torque have the same amplitude.  An estimation for the location of this zero-torque radius 
$a_S$ can be obtained 
by equating the diffusion and libration timescales across the horseshoe region. It is straightforward to show that this gives:
 \begin{equation}
 \frac{a_S}{a_0}\sim \left(40\left(\frac{\chi}{a_0^2\Omega_0}\right)^2\left(\frac{h_0}{q}\right)^3\right)^{\frac{1}{1-3f}}
 \label{eq:rstop}
 \end{equation}
 where $\Omega_0$ is the angular velocity at the initial location of the planet. For $\chi=10^{-6}$,  $q=10^{-5}$, and $f=-0.8$ we find $a_S/a_0\sim 1.6$.  Although not shown here, we find that such an estimation agrees fairly well with the value for the 
 zero-torque radius deduced from runs with the planet held on a fixed orbit. The final evolution outcome for runs with $q_d\le 0.01$ remains uncertain,  but we can reasonably expect that the migration will ultimately 
  be halted at $a\sim a_S$ for these cases.  
  For $q_d\ge 0.02$, however, the migration rate continuously increases as the evolution proceeds, leading to the planet migrating outward very rapidly. This makes the planet pass through the location of the zero-torque radius predicted by static torques and rapidly reach the outer 
  edge of the computational domain.  In that case, the migration rate is again higher than what  expected from 
  the action of the Lindblad plus the fully unsaturated entropy-related corotation torque, as illustrated by the lower right panel of Fig. \ref{fig1} which 
  shows the drift rate as a function of radius for the simulation with $q_d=0.02$. This result seems to indicate that in non-isothermal discs 
  and provided the disc mass is high enough,  a
  protoplanet can feel additional dynamical corotation torques as it migrates, similarly to what occurs 
  in isothermal discs (Paardekooper 2014).

\begin{figure}
\centering
\includegraphics[width=\columnwidth]{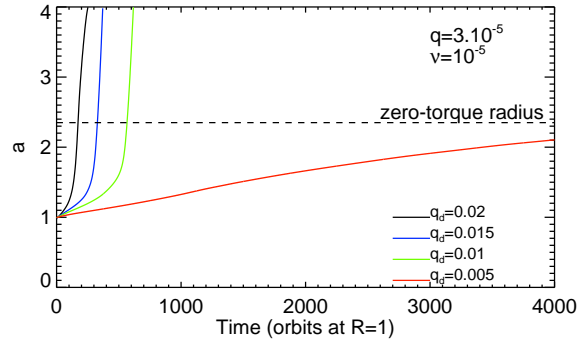}
\caption{Time evolution of the semi-major axis of a planet with $q=10^{-5}$ embedded in a non-isothermal disc with $\xi=2$ and $\nu=\chi=10^{-6}$, 
for various disc masses. }
\label{fig2}
\end{figure}

  In Fig. \ref{fig2} is presented the orbital evolution of a $q=3.10^{-5}$ planet for the same disc model, except that both the viscosity and thermal diffusivity 
have been increased to $\nu=\chi=10^{-5}$ to make sure that the underlying surface density profile is not altered by the presence of the planet. Eq. 
\ref{eq:rstop} predicts that outward migration should come to a halt at $a_S\sim 2.35$ in that case.  Although  it seems reasonable to claim 
 that migration will stall close to this location in the simulation with $q_d=0.005$, we see that the planet undergoes 
fast outward migration  and  crosses the location of the zero-torque radius in runs with $q_d\ge 0.01$.  
Here $q/h_0^3\sim 0.8$,  
so that it is plausible that here, the occurence of fast outward migration for lower disc masses might be related to the onset of non-linear effects, 
resulting in a boost of the corotation torque (Masset et al. 2006).

\subsection{Coorbital mass deficit}
\label{sec:coorb}

\begin{figure*}
\centering
\includegraphics[width=0.49\textwidth]{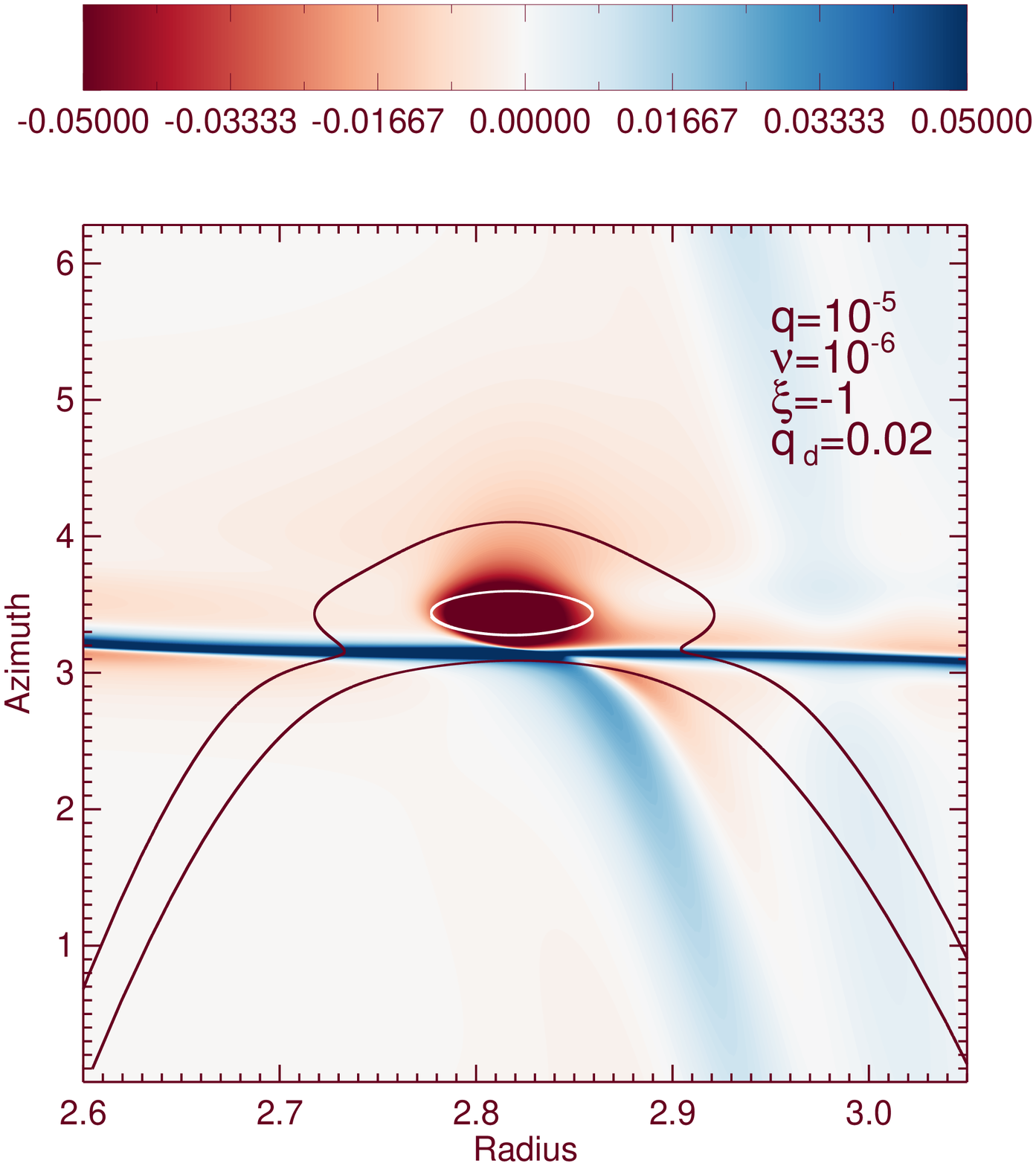}`
\includegraphics[width=0.49\textwidth]{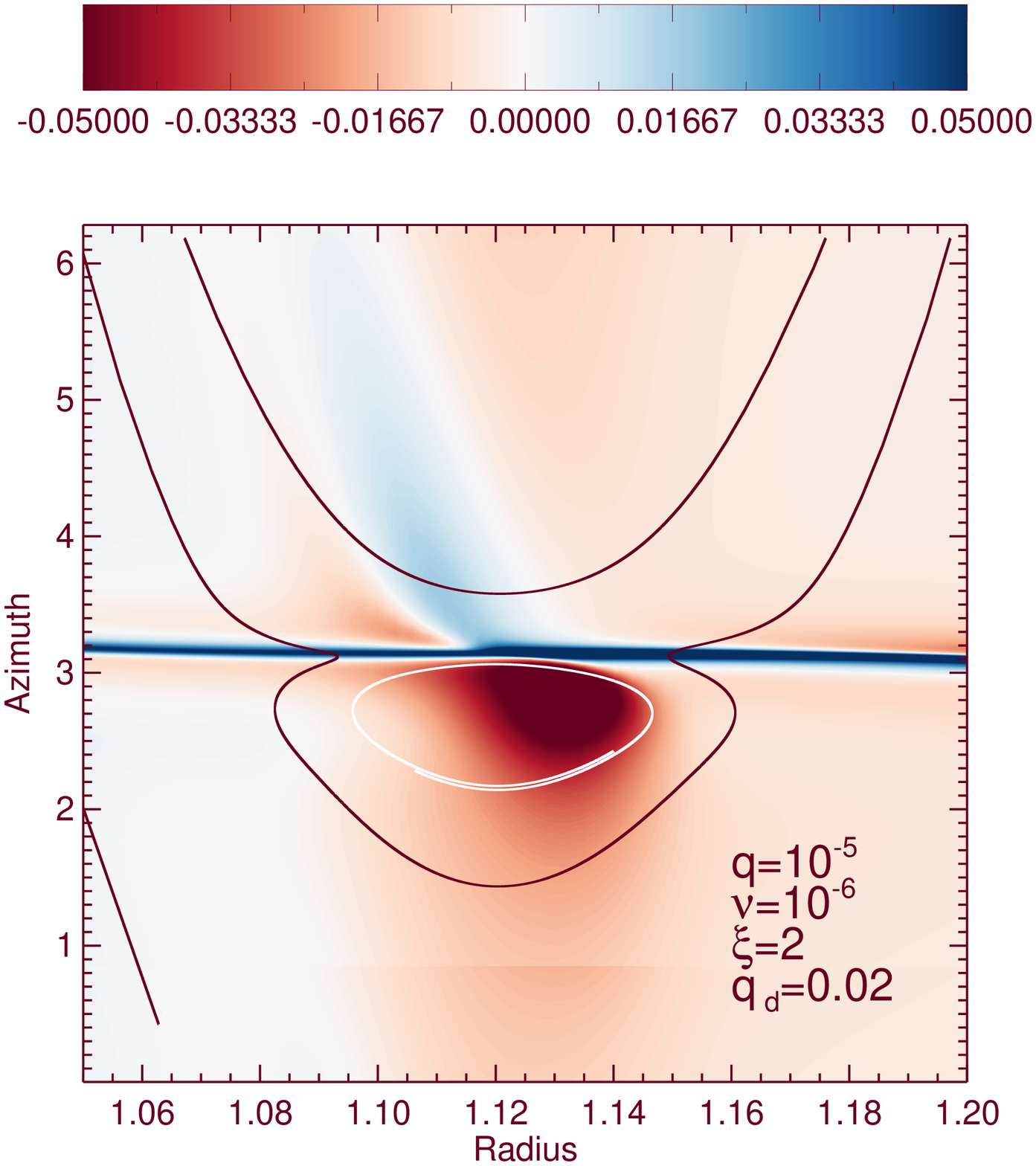}
\caption{{\it Left panel:} Contours of the perturbed surface density at $t=200$  orbits near the planet for the disc model with 
$\xi=-1$ and $q_d=0.02$. Overplotted as white lines are a few streamlines that correspond to the librating 
trapped material that 
moves with the planet and which has therefore a negative feedback on migration. The black lines are streamlines that correspond to the material that flows across the orbit as the planet 
migrates and which has a positive feedback on migration. In the case where the planet migrates in the direction set by the 
corotation torque, the trapped region is gas depleted, yielding a net positive feedback on migration.  {\it Right panel:} same but for the disc model with $\xi=2$ and $q_d=0.02$. }
\label{fig:2d}
\end{figure*}

In order to get some insight about the origin of the observed large drift rates, we display in Fig. \ref{fig:2d}  the perturbed surface density near the planet at $t=200$ orbits, for the disc model with 
$\xi=-1$ (left panel) and $\xi=2$ (right panel). For $\xi=-1$ (resp. $\xi=2$) the initial entropy gradient is positive (resp. negative) and 
dynamics in the horseshoe region yields a positive (resp. negative) surface density perturbation  at the outward downstream separatrix whereas a negative (resp. positive) surface density perturbation is created at the inward downstream separatrix. 
Overplotted as black lines are a few streamlines which correspond  to  material that flows across the orbit by executing a 
single U-turn,  whereas the overplotted white 
line shows  the librating material bound to the planet. We see that the radial drift of the planet causes the horseshoe streamlines to be significantly distorted 
and to contract into a  tadpole-like region.  This arises because the  planet drift rate $\dot{a}$ is higher than the critical drift rate $\dot{a_f}$  above which the planet migrates over a distance greater than the horseshoe 
width over one horseshoe libration time.  

As the planet migrates, the material that flows across the orbit  has always a positive feedback on migration,  whereas librating material that moves with the planet has  always a negative feedback on migration. However, in the case 
where the planet migrates in the direction set by the corotation torque, it is clear from Fig. \ref{fig:2d} that the tadpole-like region tends to be underdense compared to the rest of the disc, 
resulting in a net positive feedback on migration, and possibly leading to high drift rates. We note that this effect cannot occur if the 
planet migrates in the opposite direction to that set by the corotation torque, since in that case the librating region would be  over dense  relative to 
the ambient disc, resulting in a significant negative feedback on migration. 

  Interestingly,  the presence of such an  underdense librating region at the leading side of the planet has also been observed in simulations of giant planets which undergo inward
runaway migration (Artymowicz 2004; D'angelo \& Lubow 2008). Here, it is worth noting that the low-density region does not result from 
the gap opening process but is rather due to radiative effects. Neverthess, using the terminology employed in the 
framework of  Type III migration, Fig. \ref{fig:2d} 
shows that radiative 
effects can give rise to a coorbital mass deficit given by (Masset \& Papaloizou 2003):

\begin{equation}
 \delta m=4\pi a x_s\left(\Sigma_s-\Sigma_h\right)
 \end{equation}
 where $\Sigma_s$ is the surface density at the  upstream separatrix and $\Sigma_h$ the 
  mean surface density in the librating region. In order to provide an estimation for the coorbital mass 
  deficit in the librating region,  we make in the following the 
  assumption that the only contribution 
 to the density perturbation in the librating region results from entropy advection.  In reality, however, the density 
 perturbation features an additional contribution linked to the production of vortensity at the outgoing 
 separatrix (Paardekooper et al. 2011). Ignoring this contribution  gives the following expression 
 for  the surface density $\Sigma_-$ in the 
 librating region (Baruteau \& Masset 2008; Paardekooper et al. 2011):
 \begin{equation}
 \Sigma_-=\Sigma_s\left(1-2\frac{\mid \xi\mid  x}{\gamma a}\right) \quad \text{for} \quad  0<x<x_s
 \end{equation}
Here, we have also  ignored the radial density gradient and assumed that the corotation torque was
 initially unsaturated. This gives a mean surface density in the librating region:
 
 \begin{equation}
 \Sigma_h=\Sigma_s(1-\frac{\mid \xi \mid  x_s}{\gamma a}).
 \end{equation}
  The  coorbital mass deficit then becomes:
  \begin{equation}
   \delta m = 4\pi \mid \xi \mid x_s^2 \Sigma_s/\gamma
   \label{eq:coorb}
   \end{equation} 
which shows that the coorbital mass deficit scales with the entropy gradient and the width of the horseshoe region. 
We note in passing that at the boundary corresponding to $\dot{a} \sim \dot{a_f}$, we expect   the  corotation torque exerted on the planet 
to be  given by (e.g. Papaloizou et al. 2007):
\begin{equation}
\Gamma_{CR}=\frac{1}{2}\dot{a_f}\Omega_pa \delta m
\label{eq:gc}
\end{equation}
Using the expression for the coorbital mass deficit given by Eq. \ref{eq:coorb}, the previous equation then becomes:
\begin{equation}
\Gamma_{CR}=\frac{3}{4}\frac{\xi}{\gamma}\Sigma_s x_s^4\Omega_p^2
\label{eq:gcr}
\end{equation}
so that we recover the standard expression for the entropy-related horseshoe drag (Baruteau \& Masset 2008). For $\dot{a} > \dot{a_f}$, however,  the expression for the unsaturated corotation torque given by Eq. ~\ref{eq:gc} is no longer valid and should be modified (Papaloizou et al. 2007). 
In that case, we expect the  corotation torque to be prevented from saturation since the horseshoe region does not longer extend to the full 
$2\pi$ in azimuth, so that phase-mixing cannot occur. In isothermal discs with a large-scale vortensity 
gradient, Ogilvie \& Lubow (2006) found that drift rates
with  $\dot{a} > \dot{a_f}$ can 
even lead to torques that are much larger than the unsaturated value in absence of migration.  They showed that this arises due to corotational torques caused by a vortensity asymmetry 
in the coorbital region between gas on the leading and trailing sides of the planet. In the limit of low viscosity, the trapped material tends 
to conserve its initial vortensity as the planet migrates, leading to a possibly significant vortensity contrast between the librating 
region and the local disc.  In that case,  the impact on migration depends in fact on the sign of the vortensity deficit (Paardekooper 2014) which 
is defined  as: 

\begin{equation}
\delta(\omega/\Sigma)=1-\frac{\omega_a/\Sigma_a}{\omega_{0}/\Sigma_{0}}
\end{equation}
where $\omega_a/\Sigma_a$ is the local disc vortensity at the position of the planet and $\omega_0/\Sigma_0$ the vortensity at the initial location 
of the planet. 
In the case where the planet migrates in the direction set by the vortensity-related horseshoe drag, the vortensity deficit is positive 
and this yields a positive feedback on migration whereas if the planet migrates in the opposite direction to that set by the corotation torque, the vortensity 
deficit is negative and this gives rise to a negative feedback on migration (Paardekooper 2014).

  Here, the appearance of the 
streamlines in  Fig. \ref{fig:2d}, combined with the very high drift rates (higher than those expected from unsaturated torques)  that have been 
reported in the previous section, strongly suggests that  a mecanism similar to that presented in  Ogilvie \& Lubow (2006) is 
at work here. However, since there is no initial vortensity gradient inside the disc, we suggest  the  strong corotation torques 
acting on the planet as being generated by an entropy rather than a vortensity deficit between the trapped material and the ambient disc. For 
non-isothermal discs, we  therefore make use of 
an entropy deficit $\delta S$ that we define as:
\begin{equation}
\delta S=1-\frac{S_a}{S_0}
\end{equation}
where $S_a$ is the local disc entropy at the position of the planet and $S_0$  the disc entropy at its initial location.  In the case 
where the entropy deficit 
is positive, the planet migrates in the direction set  by the entropy-related horseshoe drag and an underdense region with an entropy excess compared 
to the local disc appears near the planet. As mentionned earlier, this yields a positive feedback on migration  and possibly gives rise to 
high drift rates if the viscosity and thermal diffusivity are small enough so that the entropy excess in the librating region is 
conserved in the course of migration. In the case where the entropy deficit 
is negative, however, the librating region  tends to be cooler and over dense relative to the ambient disc, resulting in a negative 
feedback on migration. In that case, we expect migration in non-isothermal discs to be strongly slowed down. \\

As mentionned earlier, we expect this mecanism  to arise whenever the drift timescale across the horseshoe region 
is shorter than the libration time. This condition reads (Papaloizou et al. 2007):
\begin{equation}
\dot{a} > \dot{a_f}
\label{eq:af}
\end{equation}
with (Peplinski et al. 2008):
\begin{equation}
\dot{a_f}= \frac{3x_s^2\Omega_p}{8\pi a}
\end{equation}
Using Eq. \ref{eq:af}, we can now estimate under which condition on the disc mass, horseshoe streamlines can be  significantly distorted by the effect of the 
radial drift of the planet.   The migration rate  $\dot{a}$ of the planet  is  (e.g. Paardekooper 2014):
\begin{equation}
\dot{a}=2a\frac{\Gamma/\Gamma_0}{m_p\sqrt{GM_\star a}}\Gamma_0
\end{equation}
Given that $\Gamma_0=(q/h_p)^2\Sigma_p a^4\Omega_p^2$, the previous equation can be re-written as:
\begin{equation}
\dot{a}=\frac{2}{\pi}(\Gamma/\Gamma_0)\frac{qq_d}{h_p^2}a\Omega_p
\end{equation}
Substituting the latter expression in Eq. \ref{eq:af} and using $x_s\sim1.2a\sqrt{q/h_p}$  we find that a protoplanet
may enter a  fast migration regime provided that:
\begin{equation}
\frac{q_d}{h_p}> \frac{0.27\gamma}{(\gamma\Gamma/\Gamma_0)},
\label{eq:qdsurh}
\end{equation}
 or equivalently:
\begin{equation}
{\cal Q}< \frac{3.7}{\gamma}(\gamma \Gamma/\Gamma_0),
\label{eq:criterion}
\end{equation}
where ${\cal Q}=h/q_d$ is the Toomre parameter, and where $\gamma \Gamma/ \Gamma_0$ is given by Eq. \ref{eq:gsurg0}. For 
$\gamma=1.4$ and using $\beta=\xi+(\gamma-1)\sigma$, Eq. \ref{eq:gsurg0} reads:
\begin{equation}
\gamma\Gamma/\Gamma_0=-2.5+3.9\xi-0.6\sigma
\label{eq:crit2}
\end{equation}
Not surprisingly, this predicts that it is easier for the planet to enter the fast migration regime i) for high values of the entropy gradient, and 
ii) when the planet migrates inward and the entropy gradient is positive.  This 
also shows that in the case where the planet migrates in the opposite direction to that predicted by the corotation torque, namely if $\xi$ has 
a moderate and positive value, $\gamma \Gamma/\Gamma_0$ is of the order of unity and  distortion of the 
horseshoe streamlines occurs only provided that  the disc is relatively massive. For the disc model with $\xi=-1$, we note that Eq. \ref{eq:qdsurh} predicts fast migration for $q_d\gtrsim 0.003$ which is 
 in perfect agreement with the results of the simulations. In the case with $\xi=2$, however, Eq. \ref{eq:qdsurh} predicts fast migration for $q_d\gtrsim 0.005$ whereas the right panel of Fig. \ref{fig1} shows  fast migration for  $q_d > 0.01$. This discrepancy possibly arises because the expression given by 
 Eq. \ref{eq:qdsurh} does not take into account possible saturation effects. We therefore expect this estimation to slightly underestimate the critical disc 
 mass above which fast migration should occur.  

\begin{figure}
\centering
\includegraphics[width=\columnwidth]{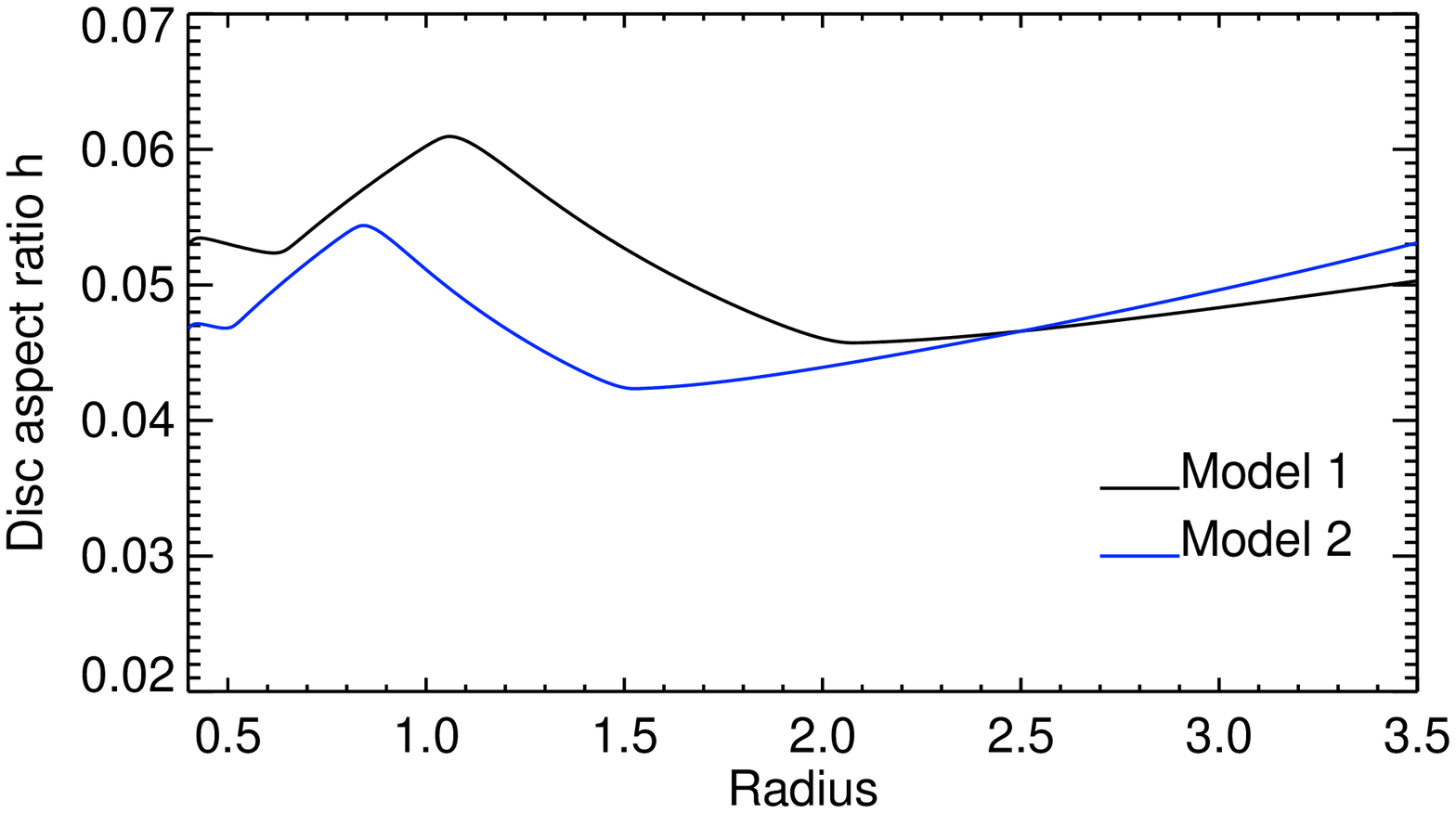}
\includegraphics[width=\columnwidth]{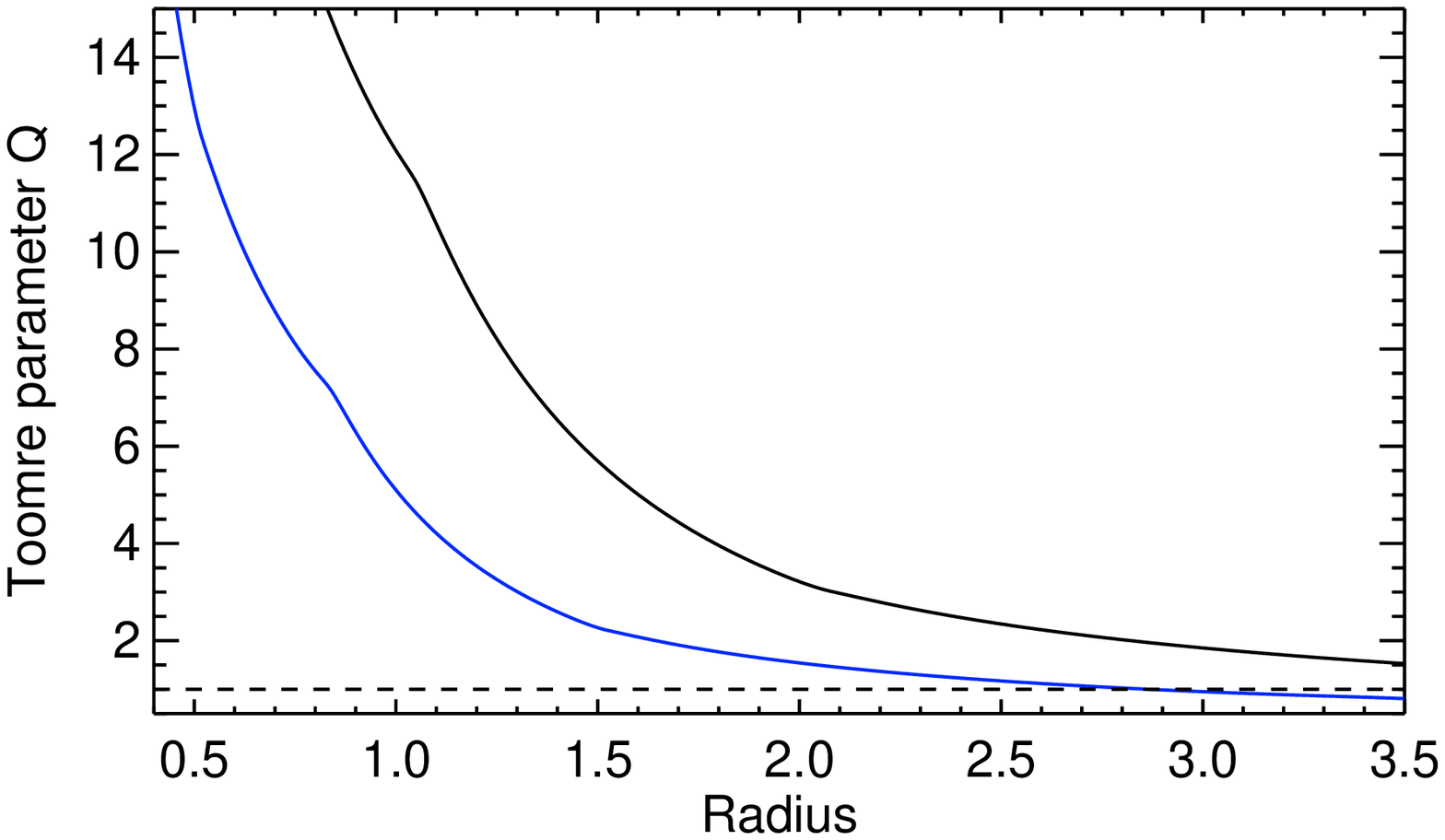}
\caption{{\it Upper panel}: Disc aspect ratio as a function of radius for the two radiative disc models that are considered and with 
parameters given in Table~$1$. {\it Lower panel}: Toomre stability parameter as a function of radius for the two models. {The dashed line 
corresponds to the gravitational stability limit.}}
\label{fig:hsr}
\end{figure}


\begin{figure*}
\centering
\includegraphics[width=0.49\textwidth]{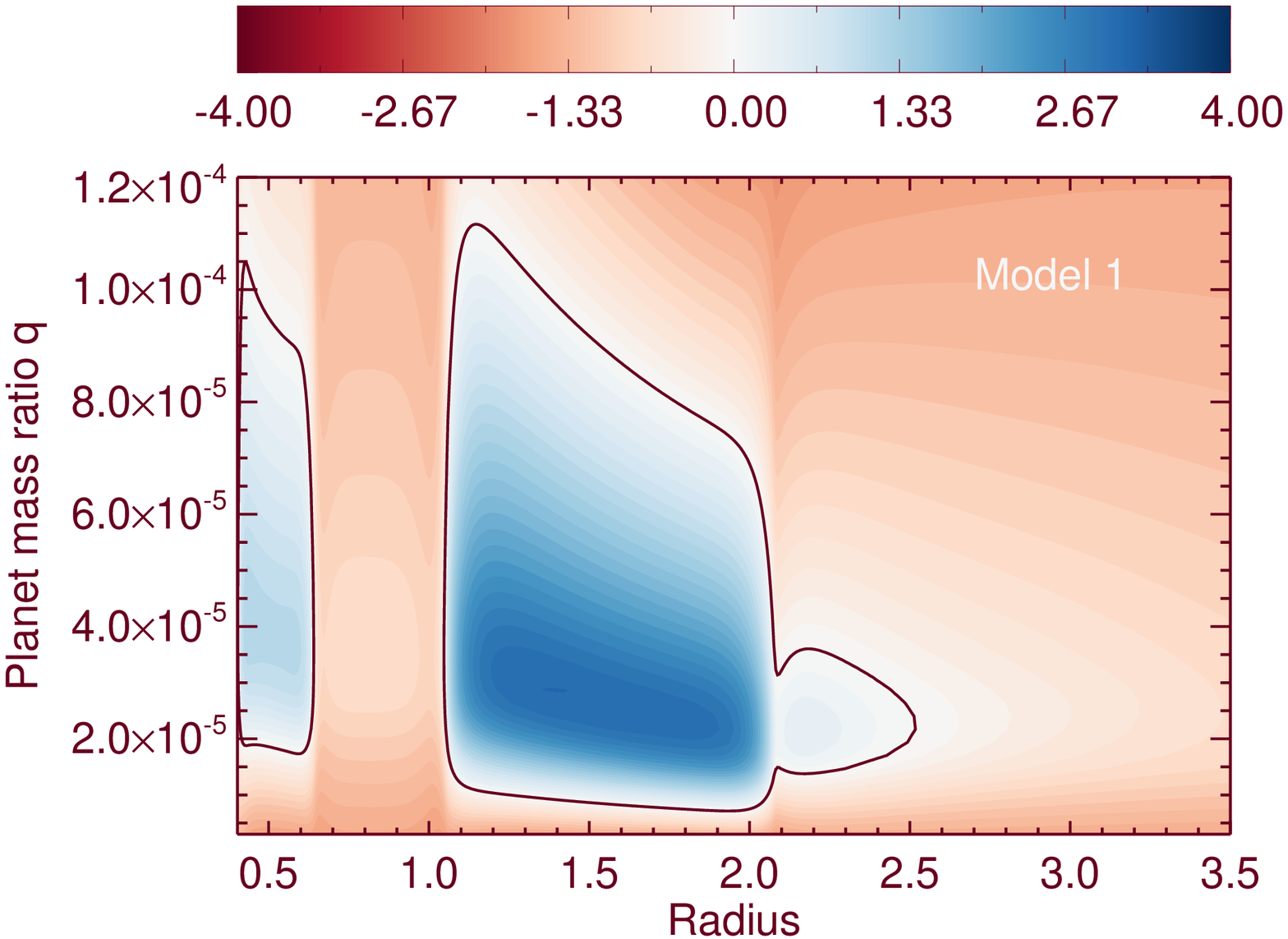}
\includegraphics[width=0.49\textwidth]{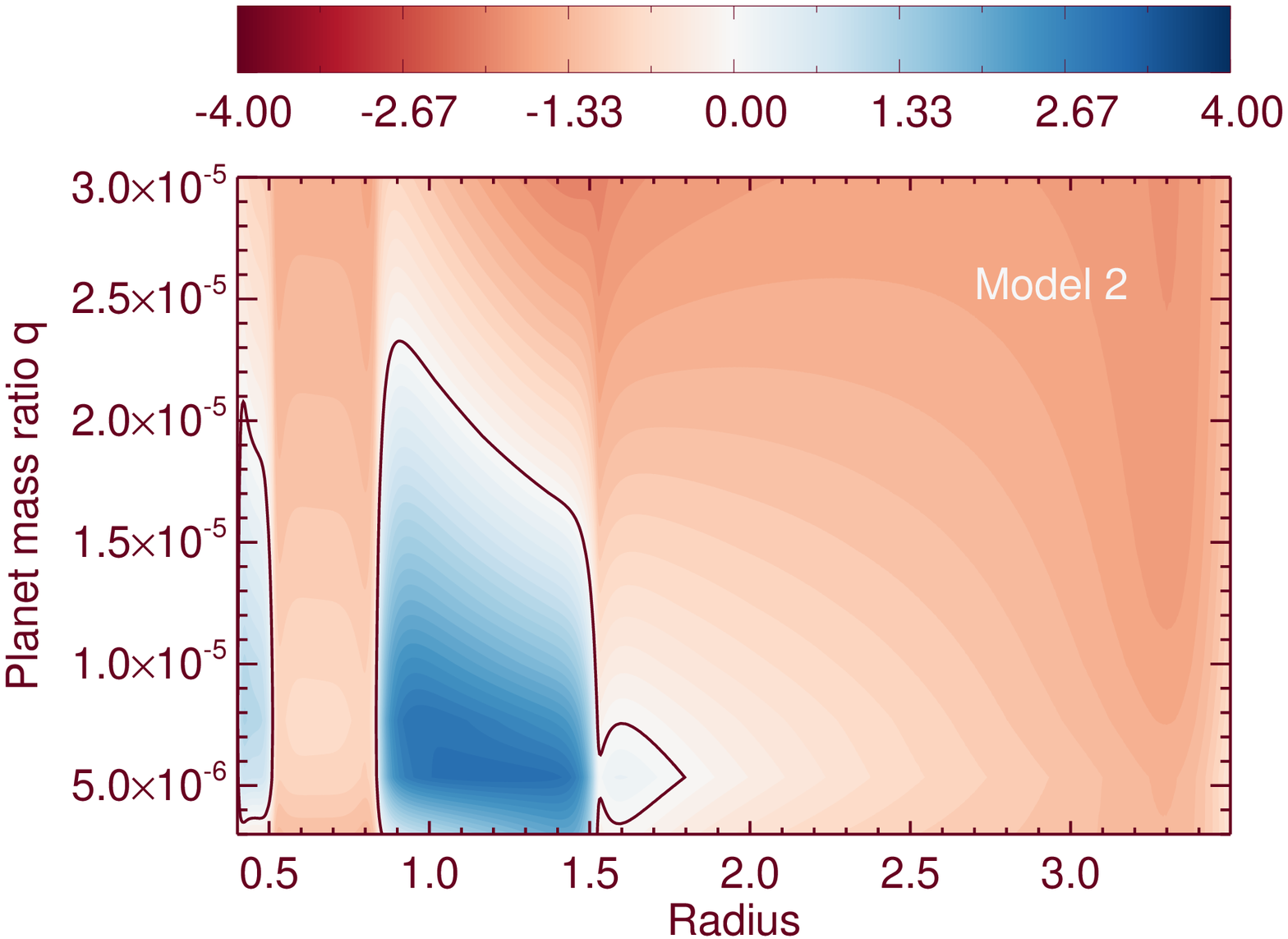}
\includegraphics[width=0.49\textwidth]{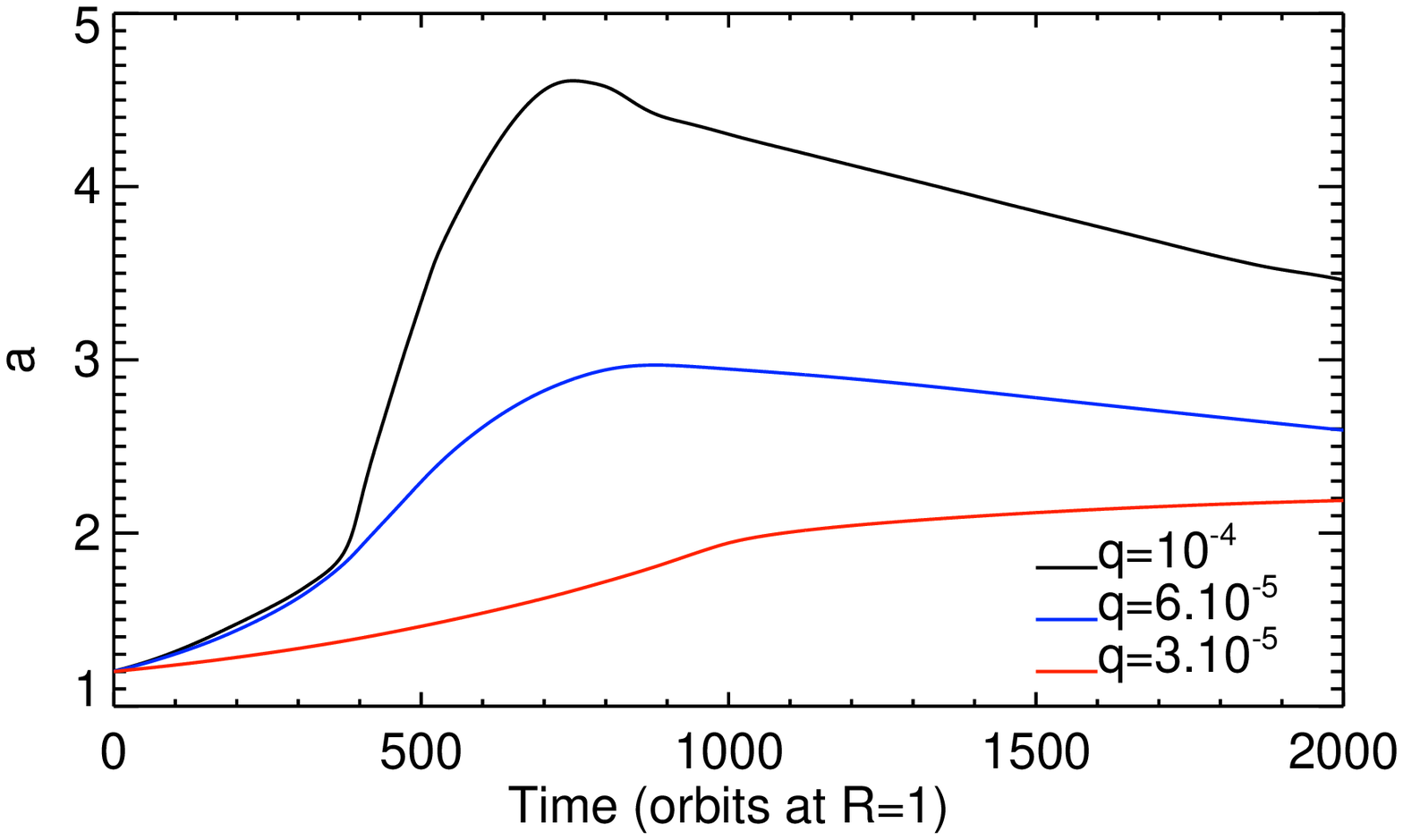}
\includegraphics[width=0.49\textwidth]{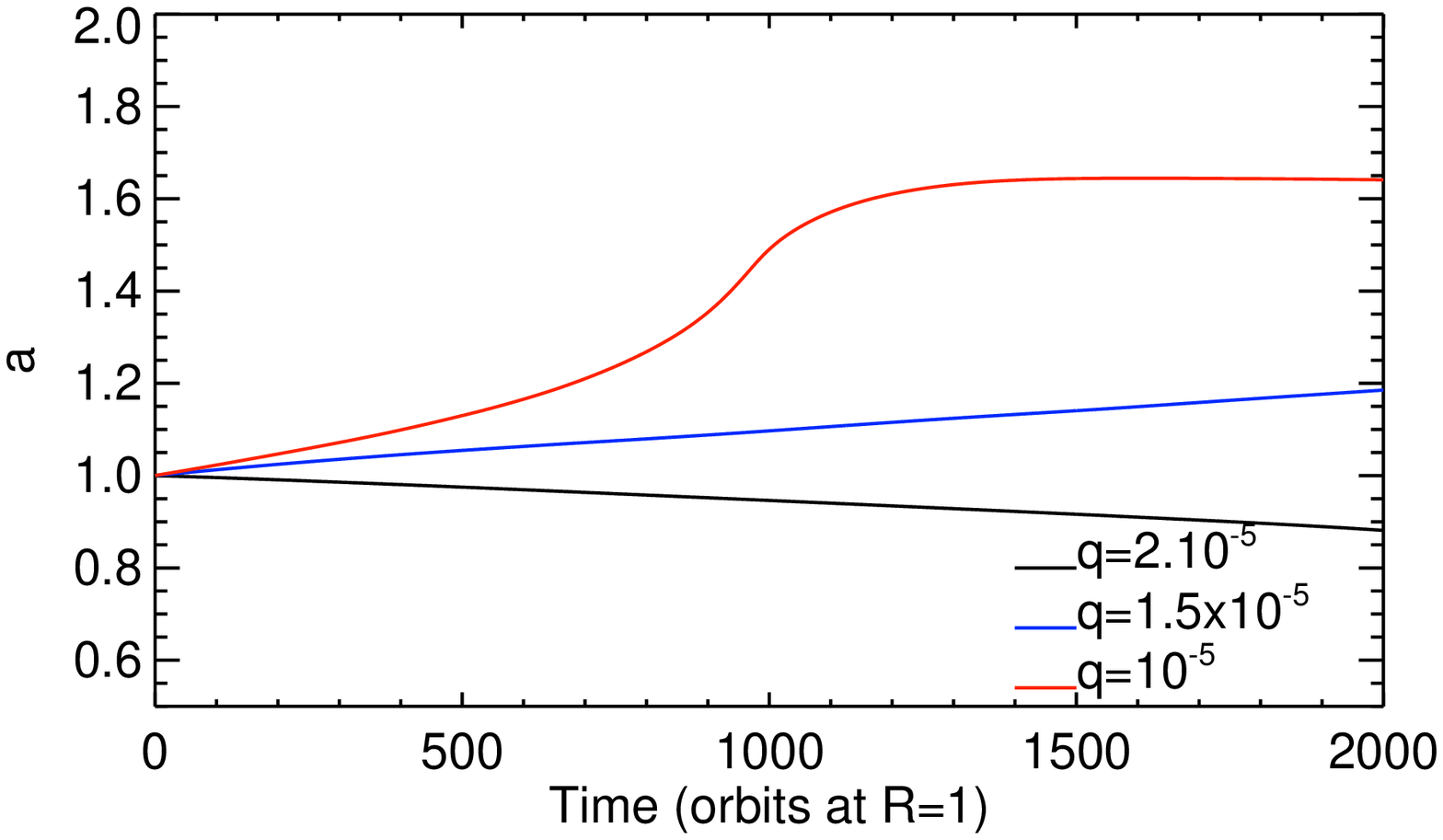}
\caption{{\it Upper panel}: Contour plots showing, for the two radiative disc models, the disc torque  as a function of planet mass and planet radial 
location. This has been derived using the analytical formulae of Paardekooper et al. (2011). {\it Lower panel:} this shows, for the two radiative discs, the time evolution of the semi-major 
axes of planets of different masses and that are initially located in the region where outward migration is expected to proceed.     }
\label{fig:maps}
\end{figure*}

\section{Radiative disc models}
\subsection{Results}

We now turn to the case of equilibrium radiative discs for which the thermodynamical state is determined by the balance between  viscous heating, stellar irradiation, and 
radiative cooling. We plot in Fig. \ref{fig:hsr} the disc aspect ratio (upper panel) and  Toomre stability parameter (lower panel) as a function 
of radius for the two models 
we consider. We remind the reader that the disc parameters for the two models can be found in Table $1$. For Model $1$, the structure in the disc aspect ratio is very similar to that of stellar equilibrium discs (SED) obtained using 3D hydrodynamical simulations (see 
for example Fig. $1$ in Lega et al.  2015). In the inner disc, viscous heating dominates over stellar irradiation and opacity transitions 
create bumps in $H/R$  at $R\sim 0.6$ and $R\sim 1$ (Bitsch et al. 2013, 2014).  In the outer parts of the disc, however, stellar irradiation is the main source of heating, leading to a flared disc with  $H/R\propto R^{2/7}$ for $R \gtrsim 2$. For Model $2$, the aspect ratio is slightly smaller 
due to the lower value for the viscosity but the radial profile of  $H/R$ looks similar to  that for Model $1$.  For both models, the torque as a function of planet 
mass and orbital distance is presented  in the upper panel of 
Fig. \ref{fig:maps}.  Following Bitsch et al. (2013), these migration maps have been obtained using the analytical formulae of Paardekooper et al. (2011) which include saturation effects for the corotation torque.  Here, the black lines delimitate  the regions in the disc where outward migration  should 
 occur. As expected, 
Type I migration is directed inward in the flared parts of the disc where the entropy gradient is positive, yielding	a negative entropy-related 
corotation torque. In the regions where the disc aspect ratio decreases with radius,  however, the entropy gradient 
tends to be negative, and this  can possibly lead to outward migration if the  corotation torque is not saturated and the entropy 
gradient steep enough.  Because the thermal diffusion timescale tends to increase with radius, outward migration will proceed in that case until the planet reaches the zero-torque 
radius where the saturated corotation torque counterbalances the Lindblad torque. As mentionned in Sect. \ref{sec:results}, this occurs when the 
thermal diffusion timescale is approximately equal to the libration timescale in the horseshoe region. Moreover, since the libration period decreases as the planet mass increases, the zero-torque radius set by  saturation tends to be located closer in  for heavier planets, which is clearly seen  in the upper panel of Fig. 
\ref{fig:maps}.  For Model $1$ (resp. Model $2$) , protoplanets with  $q\lesssim 6\times 10^{-5}$ (resp. $q\lesssim 1.6\times 10^{-5}$) would have their zero-torque radius located in the flared part of the disc 
where the entropy gradient is positive. For this mass range, the location of the zero-torque radius becomes almost mass-independent (Cossou et al. 2013) 
and rather corresponds to the transition between the viscous heating dominated regime and the stellar heating dominated region. This is located at $R\sim 2$ 
for Model $1$ while it is located at $R\sim 1.5$ for Model $2$.

We note that these migration maps have been computed from analytical formulae that are valid for planets held on a fixed circular orbit only. Returning 
to Fig. \ref{fig:hsr}, however,  we see that the Toomre stability parameter can reach values in the range $2.5<{\cal Q}<5$ in the regions where outward 
migration is expected to occur.  From the discussion presented in Sect. \ref{sec:coorb} , it  cannot be excluded that  the criterion given by Eq. \ref{eq:criterion} 
is fulfilled in this case, and that drift rates significantly higher than those predicted by static torques can be reached. More precisely, for 
both models,  the entropy gradient in the region of outward migration is estimated to be such that $\xi \sim 1.8$, which 
gives $\gamma \Gamma /\Gamma_0\sim 4$ from Eq. \ref{eq:crit2}.  The condition given by Eq. \ref{eq:criterion} therefore becomes ${\cal Q} \lesssim 10$, which is clearly fulfilled for Model $2$,  and  for Model $1$ as well  in regions with $R\gtrsim 1.3$ (see lower panel of Fig. \ref{fig:hsr}). \\

\begin{figure}
\centering
\includegraphics[width=\columnwidth]{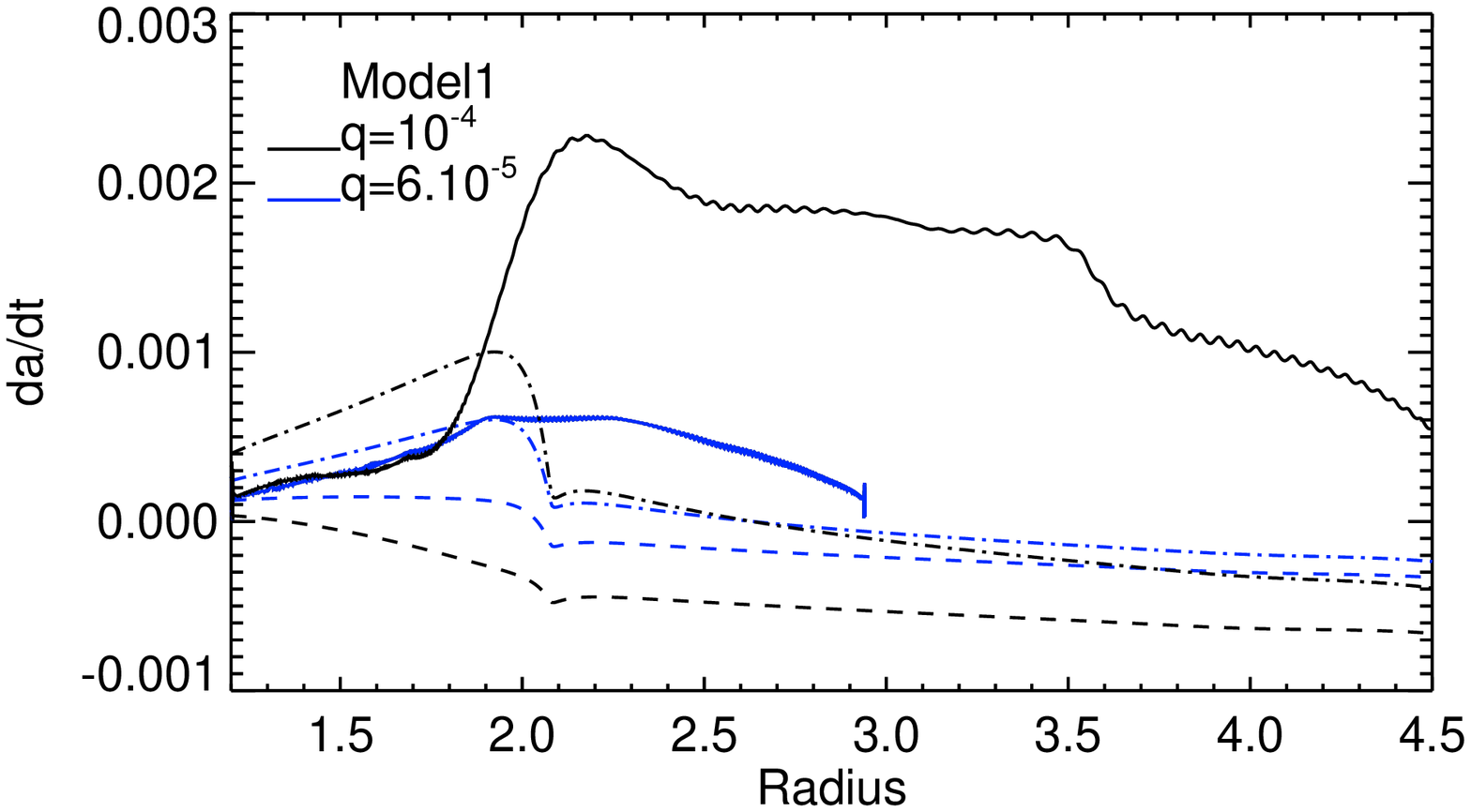}
\caption{Drift rate as a function of the  radial location of the planet for the simulations with $q=6\times10^{-5}, 10^{-4}$ and for the  radiative disc 
corresponding to Model~$1$. The dashed line represents the drift rate that is obtained using the formulae of Paardekooper et al. (2011) 
for the disc torque, whereas the dot-dashed line represents the drift rate predicted assuming a fully unsaturated 
corotation torque.}
\label{fig:da}
\end{figure}

\begin{figure*}
\centering
\includegraphics[width=0.33\textwidth]{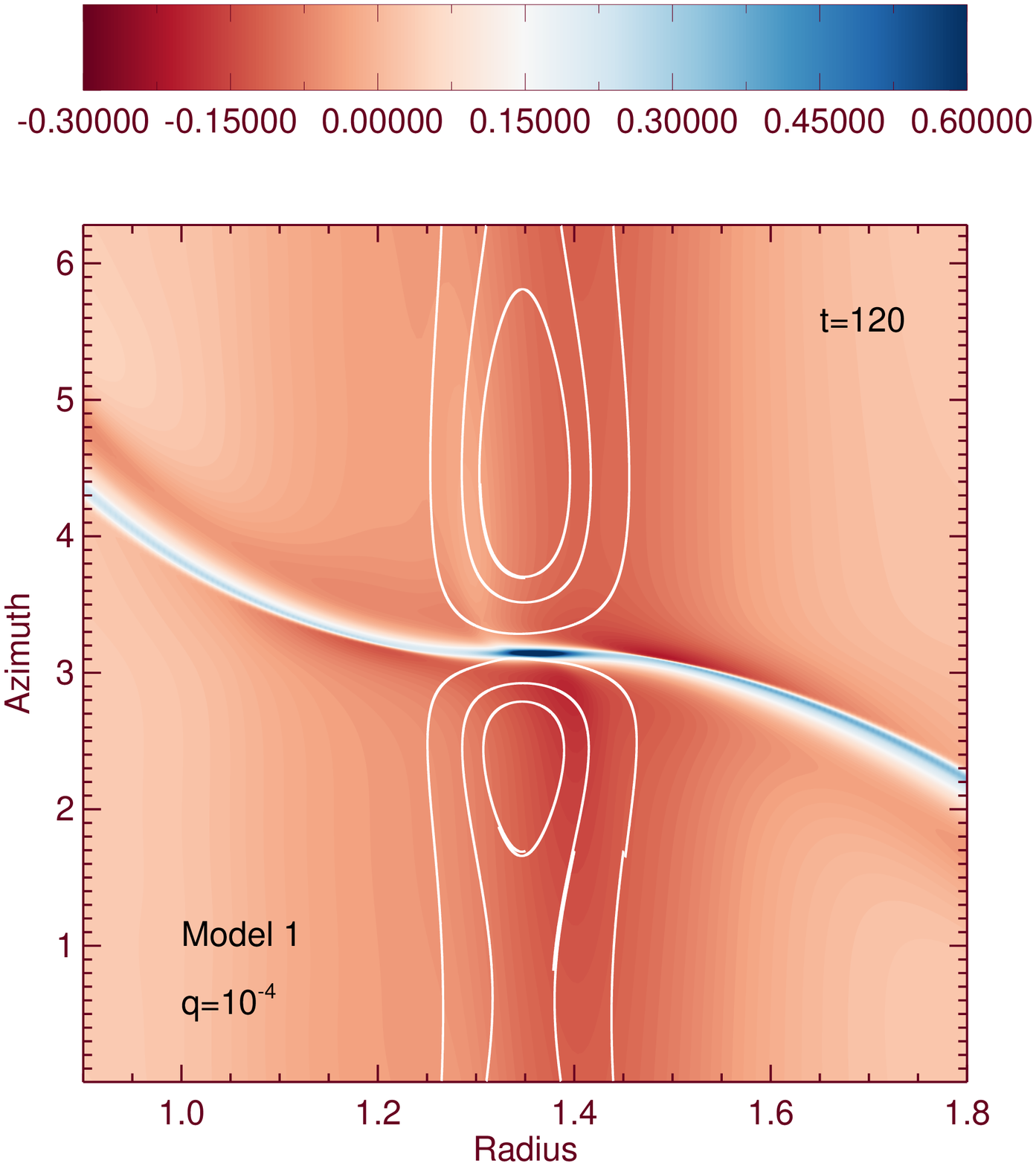}
\includegraphics[width=0.33\textwidth]{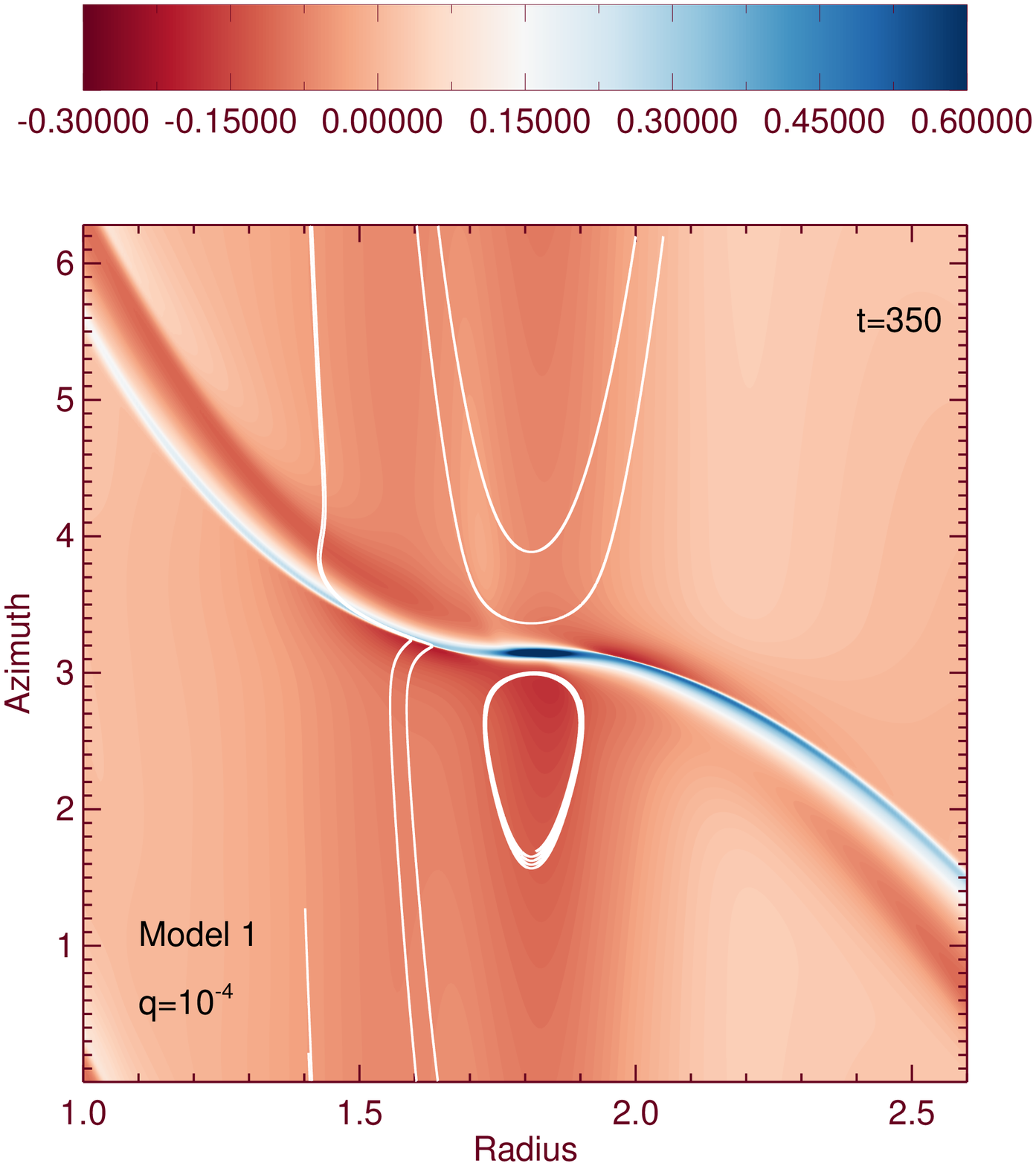}
\includegraphics[width=0.33\textwidth]{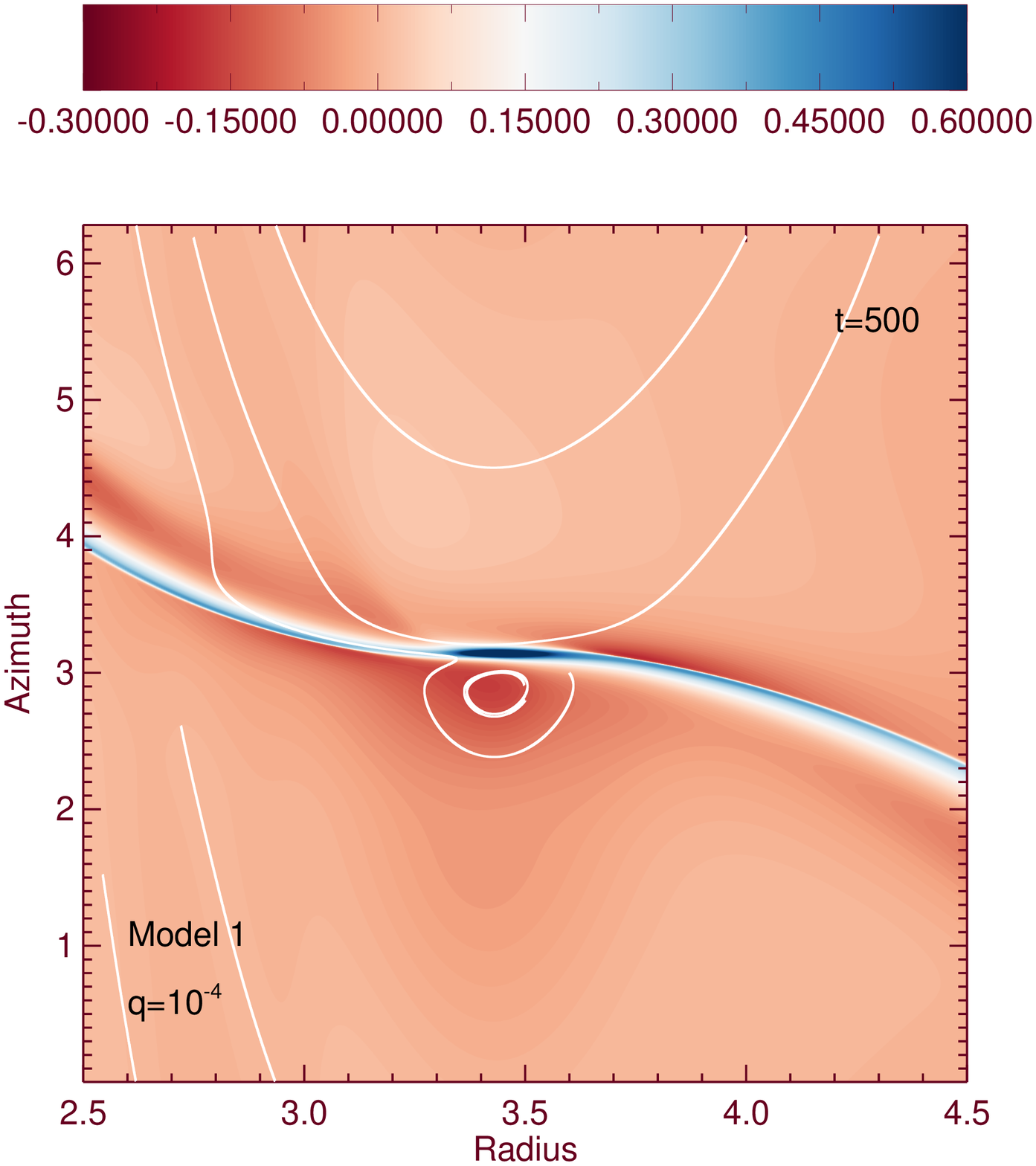}
\caption{Maps of the perturbed surface density at $t=120, 350, 500$ orbits for the simulation with $q=10^{-4}$ and for the radiative disc 
corresponding to Model~$1$. }
\label{fig:q1em4}
\end{figure*}

For Model $1$, we present in  the lower left panel of Fig. \ref{fig:maps} the semi-major axis as a function 
of time for planets with $q=3\times 10^{-5}, 6\times 10^{-5}, 
10^{-4}$ and  initially located at $a_0=1.1$, namely at a location where outward migration is expected to occur for this mass range. The orbital evolution of the 
$q=3\times 10^{-5}$ protoplanet is in close agreement with that predicted from the upper panel of 
Fig. \ref{fig:maps}, involving	 outward migration until the zero-torque radius 
located at $R\sim 2.1$ is reached. For $q=6\times 10^{-5}, 10^{-4}$, however, the evolution differs significantly to what expected from the 
migration map. The drift rates for these cases 	appear to be significantly higher than those estimated using static saturated 
or unsaturated corotation torques, 
leading the planets to migrate outward well beyond the zero-torque radius.  This is exemplified in Fig. \ref{fig:da} where we plot the planet drift rate as a function of radius for both cases. 
For the run with $q=10^{-4}$, it is clear that the planet enters a very fast migration regime once $R\sim 1.8$,  which corresponds 
to a value for the Toomre parameter of ${\cal Q}\sim 4$ (see the lower panel of  Fig. \ref{fig:hsr}).  We suspect this  value to correspond to the limit below which the mechanism presented in Sect. \ref{sec:coorb} is at work. To demonstrate that this is indeed the case, we show in Fig. \ref{fig:q1em4} a series of surface density plots 
at different times for this simulation. At early times,  there is  a negative (resp. positive) surface density perturbation at the outward (resp. inward) downstream 
separatrix due to the negative entropy gradient, and the horseshoe region is clearly visible.  At $t\sim 350$ orbits, namely just before the planet 
starts its fast outward migration, an underdense librating region has appeared, suggesting thereby that the condition given by Eq. \ref{eq:af} is 
verified at that time. As mentionned in Sect. \ref{sec:coorb}, this can yield a strong 
positive corotation torque which is responsible for the high drift rates observed. As can be seen in the third panel of 
Fig. \ref{fig:q1em4}, the trapped region progressively shrinks in the course of migration and a part of  the coorbital mass deficit can be lost, resulting in the planet migrating inward again 
at later times. Interestingly, we note that a similar process prevents outward runaway migration (Masset \& Papaloizou 2003) to be sustained
over long timescales.  Here, thermal diffusion acts also against fast migration since it makes the trapped region cool down, leading to a slight increase in the surface 
density.  Once the planet migrates inward again and evolves in the flared part of the disc, the thermodynamical state of the disc is close to the isothermal limit so that 
radiative effects can not lead to the formation of an underdense librating region near the planet, resulting in the planet migrating under the 
action of static torques only.   \\

\begin{figure}
\centering
\includegraphics[width=\columnwidth]{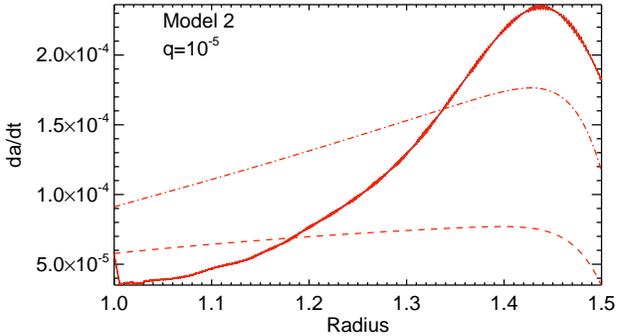}
\caption{Drift rate as a function of the  radial location of the planet for the simulations with $q=10^{-5}$ and for the  radiative disc 
corresponding to Model~$2$. The dashed line represents the drift rate that is obtained using the formulae of Paardekooper et al. (2011) 
for the disc torque, whereas the dot-dashed line represents the drift rate predicted assuming a fully unsaturated 
corotation torque.}
\label{fig:da1}
\end{figure}

For Model $2$, the orbital evolution of planets with $q=10^{-5}, 1.5\times 10^{-5}, 2 \times 10^{-5}$ is shown in the lower right panel of 
Fig. \ref{fig:maps}. Outward migration here  occurs for $q\le 1.5 \times 10^{-5}$ whereas saturation of the corotation torque makes the 
$q=2\times 10^{-5}$ protoplanet migrate inward. In the latter case, we note that this does not agree with the migration map in the upper panel of Fig. \ref{fig:maps} which predicts outward migration for $R\lesssim 1.2$. Inspection of this migration map also reveals that compared to 
the case with $q=1.5\times10^{-5}$, the level of saturation is clearly initially smaller for a planet mass with $q=10^{-5}$.  The resulting 
faster outward migration makes the $q=10^{-5}$ body rapidly reach regions with  relatively low ${\cal Q}$ values. Analysis of the streamlines reveals that 
the horseshoe region is destroyed at $t\sim 700$ orbits, at which time the planet has migrated to a disc region  where  ${\cal Q}\sim 3.5$. Again, this 
is fairly consistent with the estimation given by Eq. \ref{eq:criterion}. From that time, the planet undergoes an episode of 
fast migration due to the 
effect presented above, but becomes ultimately trapped at the zero-torque radius located at the transition between the 
viscous heating dominated and stellar heating dominated regimes. In agreement with the discussion of Sect. \ref{sec:coorb}, we find that the typical drift rates
during fast migration can be higher than those predicted assuming a fully unsaturated corotation torque, as illustrated by Fig. \ref{fig:da1} which 
shows for this run the drift rate of the planet as a function of its radial location in the disc. 

 For $q=6\times 10^{-5}$, we mention that continuation of the run indicates that the planet does not  appear to experience fast migration, in spite of 
 the fact that it can evolve in disc regions  with  ${\cal Q}\sim 3.5$. This possibly arises because, as mentionned above, the corotation torque 
 is  quite saturated in that case, resulting in a low contrast of entropy initially between the trapped material and the local disc.

\subsection{Implication for the formation of giant planet cores}
\label{sec:4planets}
It has been suggested that the zero-torque radius may represent an ideal site for the growth of giant planet cores through giant impacts 
between Earth-mass embryos (Lyra et al. 2010; Hasegawa \& Pudritz 2011). Pierens et al. (2013) have studied the evolution of multiple 
planets of a few Earth masses which migrate convergently toward a convergence zone located at the transition between two 
different opacity regimes. They found that the embryos tend to form a stable resonant chain that protect them from close encounters 
with other bodies, preventing thereby the subsequent formation of giant planet cores.  In the case of a large number of initial embryos or if the effects of disc turbulence are included these resonant configurations can be 
disrupted, but the formation of massive cores remains marginal. In a subsequent study,  
Zhang et al. (2014) focused on the case where the zero-torque radius is located at the transition between the inner viscously heated part 
of the disc and the outer irradiated region. They found that in massive discs with accretion rates $\dot M \sim 10^{-7} M_\odot/yr$ the resonant barriers that 
are formed can be broken, enabling the formation of massive cores.

In massive discs, the mechanism presented in Sect. \ref{sec:coorb} may also facilitate the disruption of resonances at convergence zones and 
consequently promote collisions between embryos.  To investigate this issue in more details, we have examined the evolution of four planets 
with $q=10^{-5}$ embedded in the disc corresponding to Model $2$, and  located in the vicinity of the zero-torque radius at $R\sim 1.5$. According 
to the results presented in the previous section, the bodies located inside the zero-torque radius are expected to rapidly 
migrate outward. \\
Here, we used a setup similar to that presented in Pierens et al. (2013), with two embryos initially located on each side of the zero-torque radius and  separated by $4.5\;R_{mH}$, where $R_{mH}$ is the mutual Hill radius.  The protoplanets are initially set on moderately inclined 
orbits, and a prescription for the rate of inclination damping due to the interaction with the disc is employed (see Pierens et al. 2013).

\begin{figure}
\centering
\includegraphics[width=\columnwidth]{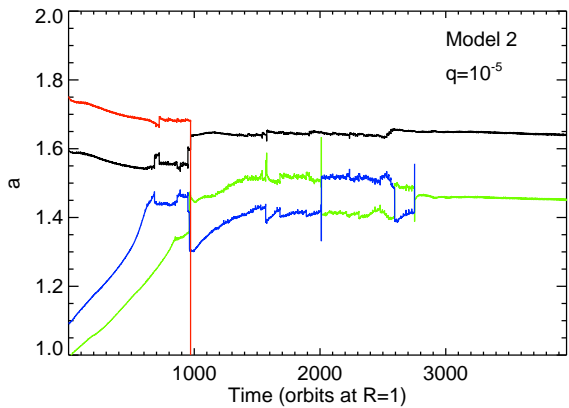}
\caption{Orbital evolution of four embryos with mass ratio $q=10^{-5}$ initially located near the transition between the viscous heating dominated 
and stellar heating dominated regimes, for the radiative disc model corresponding to model $2$.}
\label{fig:4planets}
\end{figure}

 Fig. \ref{fig:4planets} shows the time evolution of the semi major axis 
for the $4$ protoplanets. At early times, the two innermost (resp. outermost) bodies located inside the zero-torque radius undergo outward (resp. 
inward) migration, as expected. For the two innermost bodies, fast outward migration is observed, in agreement with the results of the 
previous section.  This causes the embryo initially located at $a_0=1.1$ (blue) to rapidly reach the zero-torque radius and to enter in a $9:8$ resonance 
with the third planet (black) at $t\sim 700$ orbits. The inward migrating fourth planet (red) then catches up with these two bodies and enter in a $9:8$ 
resonance with them. This resonant chain is however subsequently destabilized due to the fast outward migration of the first body (green), resulting 
in a physical collision between the two outermost planets at $t\sim 1000$ orbits. The  $6.6$ $M_\oplus$ planet  that is formed at that time by this 
process subsequently migrates inward until it reaches the zero-torque 
radius at $R\sim 1.5$. Regarding the  two innermost embryos, they  undergo a series of 
orbital exchanges inside the zero-torque until they collide at $t\sim 2800$ orbits, leading again to the formation of a $6.6$ $M_\oplus$ planet. From the migration map corresponding to model $2$,  we expect this body to slowly migrate inward at later times until it reaches the zero-torque  radius 
set by saturation and which is located at $R\sim 1.2$ for such a planet mass.\\
 This result seems to indicate that in relatively massive discs for which the condition given by Eq. \ref{eq:criterion} is fulfilled, fast convergent migration 
 of low-mass embryos may indeed enhance the formation of massive cores through collisional growth of low-mass bodies. We will address in a future publication the
  issue of the dependence of these results on the 
 disc model and  initial mass of the embryos, and investigate whether or not giant planet cores may indeed be formed by this process.

\section{Discussion and conclusion}

In this paper, we have presented the results of hydrodynamic simulations of the orbital evolution of embedded low-mass planets in 
non-isothermal and radiative discs.  A main aim of this study is to  examine the dependence of the planet drift rate on the disc 
mass and entropy gradient inside the disc, with particular emphasis put on the possible role of dynamical corotation torques. We consider non-barotropic discs with an initial radial entropy gradient, and in which both the thermal diffusion coefficient and viscosity are constant. We also investigate the case of  radiative disc models that 
include  the effect of viscous heating, stellar irradiation and radiative cooling.
For non-isothermal disc models with an initial non-zero entropy gradient and zero vortensity gradient, we find that the drift rates of a migrating protoplanet can be, under certain conditions, much 
higher than those expected from classical formulae for the disc torque. We observe that these can even be higher than those expected 
assuming a fully unsaturated horseshoe drag. This arises provided that  i) the Toomre stability parameter is less than a critical value  that depends 
on the entropy gradient inside the disc and which is given by Eq. \ref{eq:criterion} and 
ii) the planet migrates in the direction set by the entropy-related horseshoe drag.  If condition i) is satisfied, the horseshoe region of the planet 
does not extend to the full  $2\pi$ in azimuth and rather contracts into a tadpole-like region (Papaloizou et al. 2007). This region is bound 
to the planet and has therefore  
a negative feedback on migration. It is  located at the leading side  of the planet if the latter migrates  inward,  whereas 
it is located at trailing side  of the planet if migration is inward.
However, if condition ii) is also verified, the trapped material shows an entropy excess and becomes underdense relative to the ambient disc 
 as the disc tries to maintain a pressure balance. In that case, the main contribution to the 
corotation torque comes from the material that flows across the orbit as the planet migrates and which yields a positive feedback on migration. This can 
make the planet enter in a fast migration regime, provided that the viscosity and thermal diffusivity are small enough so that the entropy 
excess in the trapped region tends to be conserved in the course of migration.

As a side result, we  find that an outward migrating planet that  enters such a fast migration mode can pass through the location of the zero-torque radius 
due to saturation effects, in agreement with the results of Paardekooper (2014) in isothermal discs.

Our simulations for radiative disc models essentially confirm these findings. For a stellar equilibrium disc with constant viscosity $\nu=10^{-5}$ and disc mass equivalent to 
$5$ times the Minimum Mass Solar Nebula (MMSN) at 5 AU, planets with mass ratio in the range $6\times 10^{-5}<q<10^{-4}$ are 
observed to enter a fast migration regime in the region $5<R<10$ AU where outward migration should proceed according to the 
corresponding migration map (see Fig. \ref{fig:maps}). In agreement with the analytical estimation given by Eq. \ref{eq:criterion},   this occurs 
once the Toomre stability parameter becomes smaller than the critical value ${\cal Q}\sim 3-4$.  From that time,  strong corotation torques
make the planet not only pass through the location of the zero-torque radius set by saturation, but also migrate well beyond the boundary 
corresponding to  the transition between the  viscous heating dominated regime and the stellar heating dominated regime. For 
example, we observe that fast outward migration of Neptune-mass planets 
can proceed until the planet reaches  $R\sim 20$ AU , whereas  migration contours predict inward migration  outside $~7$ AU. 
For a model with  $\nu=10^{-6}$ and disc mass equivalent to 
$2.5$ times the MMSN at 5 AU, we find that the mechanism presented above can strongly affect the outward migration of protoplanets
with $q\sim 10^{-5}$ (equivalent to $3.3\;M_\oplus$). In that case, it appears that the drift rate is again higher than the one expected assuming a fully unsaturated corotation torque, but the planet becomes ultimately trapped at the transition between the viscous heating dominated and stellar 
heating dominated regimes.  
For $q=10^{-5}$, we have examined the impact of this fast migration regime on the orbital evolution of multiple planets that convergently 
migrate at this transition. We find that the high drift rates that are achieved in the course of outward migration may help in disrupting the resonant chains that are established between embryos. The consequence is that the process of giant core growth by collisions of low-mass bodies may be easier if 
these embryos undergo fast migration, but this needs to be investigated in more details.

A major simplification of this work resides in the fact that we have not considered the  effect disc self-gravity.   However, 
we have seen that the typical value for the Toomre stability parameter below which fast migration is triggered is typically ${\cal Q}\sim 3-4$ in radiative 
discs. For such a value, the shift in the positions of the  Lindblad resonances caused by self-gravity (Pierens \& Hur\'e 2005) is expected 
to lead to a Lindblad torque which is $\sim 15 \%$ stronger in comparison with the case where self-gravity is discarded (Baruteau \& Masset 2008). 
In the case where the planet migrates outward (resp. inward), we 
therefore expect  Eq. \ref{eq:criterion} to slightly underestimate (resp. overestimate)  the value for ${\cal Q}$ below which the horseshoe 
streamlines become significantly distorted by the radial drift of the planet.  Clearly, additional self-gravitating simulations  are required to 
definitely assess how these results are impacted by including the effect of the disc self-gravity.  This issue will be examined in a future 
publication.

\section*{Acknowledgments}
We thank A. Morbidelli, R. Nelson and S. Raymond for useful discussions. Computer time for this study was provided by the computing facilities MCIA (M\'esocentre de Calcul Intensif Aquitain) of the Universite de Bordeaux and by HPC resources of Cines under the allocation c2015046957 made by GENCI (Grand Equipement National de Calcul Intensif).  We thank the Agence Nationale pour la Recherche under grant ANR-13-BS05-0003 (MOJO).

\end{document}